\documentclass[journal]{IEEEtran}

\usepackage{ragged2e}
\usepackage{floatrow}
\usepackage{amsthm}
\usepackage[utf8]{inputenc}
\usepackage[english]{babel}
\usepackage{multicol}
\usepackage{epsfig}
\usepackage{epstopdf}
\usepackage{float}
\usepackage{perpage} 
\usepackage{array}
\usepackage{graphicx}
\usepackage{amsfonts}
\usepackage{comment}
\usepackage{subcaption} 
\usepackage{algorithm}
\usepackage{algpseudocode} 
\MakeSorted{figure}
\MakeSorted{table}
\usepackage{bbm}
\usepackage{bm,xstring}
\usepackage{xcolor}
\usepackage{multirow}
\usepackage{soul}
\usepackage{cite}
\usepackage{array} 
\usepackage[utf8]{inputenc}
\usepackage{amsmath}
\usepackage{amssymb}
\theoremstyle{remark}
\newtheorem{theorem}{Theorem}

\usepackage{tabularx}
\usepackage{pifont} 

\let\labelindent\relax
\usepackage{enumitem}
\usepackage{algpseudocode}  
\usepackage{hyperref}
\usepackage{cite}
\usepackage{fixltx2e}
\usepackage[font=footnotesize,labelfont=bf, figurename=Fig.]{caption} 
\usepackage{fix-cm}
\usepackage{array, makecell} 
\usepackage{tabularx}
\usepackage{multirow}
\usepackage{makecell}

\begin{document}

\title{SFiD: Self-Supervised Fitness-Driven Optimization for Next-Generation Networking}
\author{Kalpesh K. Patel, \textit{Graduate Student Member, IEEE},  Ekant Sharma, \textit{Senior Member, IEEE}, \\ Sandeep Kumar Singh, {\textit{Senior Member, IEEE}} and  Lajos Hanzo, {\textit{Life Fellow, IEEE}}
\thanks{Kalpesh K. Patel, Ekant Sharma and Sandeep Kumar Singh are with the Department of Electronics and Communication Engineering, Indian Institute of Technology Roorkee, India (e-mail: \{kalpeshkumar\_p, ekant, sandeep.singh\}@ece.iitr.ac.in).}
\thanks{Lajos Hanzo is with the Department of Electronics and Computer Science, University of Southampton, SO17 1BJ Southampton, U.K. (e-mail:
lh@ecs.soton.ac.uk)}
\vspace{-30pt} }
\maketitle

\begin{abstract}
Next-generation (NG) wireless networks face increasingly complex and dynamic optimization challenges driven by data-intensive applications and massive device connectivity. Existing methods suffer from one or more limitations, including high computational complexity, poor scalability and adaptability, reliance on large labeled datasets, or weak generalization in dynamic and non-convex wireless environments. To overcome these limitations, we introduce a self-supervised fitness-driven diffusion (SFiD) method that autonomously generates its own training data and iteratively learns from it. It also exhibits strong generalization capabilities and the ability to effectively tackle challenging optimization problems. We validate the convergence of our proposed method to the global optimum on both analytical convex and non-convex optimization problems. To demonstrate the effectiveness of the proposed method, we consider reconfigurable intelligent surface (RIS) phase-shift optimization in cell-free massive multiple-input multiple-output (CF-mMIMO) systems with imperfect channel state information and spatial correlation at the RIS as a case study. Specifically, we develop SFiD-C for continuous variable optimization and apply it to optimize continuous RIS phase shifts in CF-mMIMO systems. We also develop SFiD-D for discrete variable optimization and apply it to optimize discrete RIS phase shifts in CF-mMIMO systems. 
Furthermore, we develop a binary neural network (BNN) based version of SFiD-D for reducing its computational complexity, referred to as SFiD-D (BNN). Finally, our experimental results demonstrate that the proposed methods markedly improve computational efficiency and achieve comparable or superior performance compared to meta-heuristic and DRL benchmarks for both continuous and discrete RIS phase-shift optimization.
\end{abstract}

\IEEEpeerreviewmaketitle
\vspace{-10 pt}
\section{{Introduction}}
The rapid advances in artificial intelligence (AI) have been predominantly driven by the emergence of Generative AI (GAI)~\cite{In_001_GAI_2025, In_002_DMbeatGAN, In_003_GAI_2025,  In_004_DM_2025, In_005_GAI_2024}. GAI is effective in optimization and generating new content by learning patterns from existing datasets~\cite{In_003_GAI_2025, In_004_DM_2025, In_006_DiffSG_2025}. This allows GAI models to generate new data that reflects the patterns and features of the training data, while avoiding exact replication. GAI relies on a sophisticated blend of models and techniques that form the foundational framework supporting the evolution of the modern GAI ecosystem. Notable examples include variational autoencoders (VAEs)~\cite{In_007_VAE_2025}, generative adversarial networks (GANs)~\cite{In_008_GAN_2018}, and diffusion models~\cite{In_009_DDPM}, each playing a significant role in advancing AI-driven content generation and optimization.

Diffusion models offer several advantages over GANs and VAEs~\cite{In_002_DMbeatGAN, In_003_GAI_2025}. In contrast to GANs, which suffer from mode-collapse and training instability, diffusion models provide stable training through an iterative denoising process. In contrast to VAEs, which often produce low quality results due to constraints imposed by latent space regularization, diffusion models generate high-quality outputs with improved mode coverage and detail. By capturing fine-grained distributional information, diffusion models are capable of producing content rich in detail and diversity, while effectively mitigating mode-collapse. The incorporation of guidance mechanisms further strengthens diffusion models by enhancing controllability, increasing efficiency, while producing accurate outputs in conditional generation tasks~\cite{ In_011_Cond_DM_2025, In_012_CDDPM_letafati_2025, In_013_Cond_CDDM_2025}.

Optimization plays a fundamental role in next-generation (NG) wireless systems by facilitating the efficient allocation of limited resources to achieve desired performance objectives under given constraints. However, NG wireless systems face critical challenges, such as supporting massive data-driven applications and accommodating an escalating number of user equipment (UE).  These NG systems result in unique optimization problems underlying system design, posing significant challenges in their analysis and solution. These challenges include, but are not limited to, optimizing resource allocation, maximizing achievable rates, and enhancing energy efficiency or their joint multi-component Pareto optimization~\cite{In_014_Pareto_2020}. 
These NG wireless systems exhibit high complexity and dynamic behavior that require agile adaptation to rapidly fluctuating conditions, where efficient optimization plays a vital role across diverse wireless communication scenarios~\cite{In_001_GAI_2025}. Consequently, the resultant optimization problems are often highly irregular, characterized by nonconvexity, nonsmoothness, nonseparability, and have uncertainty with design variables spanning continuous, discrete, and mixed domains. These properties lead to a highly complex feasible solution space, making such optimization problems particularly challenging to solve.

\begin{table*}[!t]
\centering
\caption{Comparison of optimization approaches with its advantages and limitations.}
\vspace{-10 pt}
\label{Table-1}

\renewcommand{\arraystretch}{1.2}

\begin{tabular}
{|>{\raggedright\arraybackslash}p{3.6cm}|>{\raggedright\arraybackslash}p{3.5cm}|p{7.9cm}|>{\raggedright\arraybackslash}p{1.5cm}|}
\hline
 \textbf{Optimization Approach} & \textbf{Name~/~type of method} & \textbf{Key Advantages~/~Limitations} & \textbf{Ref.} \\ 
\hline

\multirow{2}{=}{Meta-heuristic optimization}
 & Genetic algorithm (GA) 
 & \multirow{2}{=}{Capable of producing near-optimal / optimal solutions at high computational burden and slow convergence.}
 & ~\cite{In_016_Rate-Splitting_2025, IN_017_GA_Power_2022} \\ \cline{2-2}\cline{4-4}
 & Particle swarm optimization (PSO) 
 & 
 & \cite{In_018_PSO_Malay_2025} \\
\hline

\multirow{2}{=}{Conventional optimization}
 & Successive convex approximation (SCA) 
 & \multirow{3}{=}{Address non-convex problems through tractable sub- problems but converge to local optima, \,sensitive to initialization \, and \, scale poorly in high dimensional settings.}
 & \cite{In_019_EE_BCD_SCA_2023, In_020_BCD_2021, In_021_SCA_Phase_2025, In_022_SCA_2021} \\ \cline{2-2}\cline{4-4}
 & Block coordinate descent (BCD)
 & 
 & \cite{In_019_EE_BCD_SCA_2023, In_020_BCD_2021,  In_021_SCA_Phase_2025} \\
\hline

\multirow{2}{=}{Machine learning-based optimization}
 & Discriminative model 
 & Computationally efficient but requires large labeled datasets. It relies on fixed mappings and generalizes poorly to dynamic or highly non-convex settings.
 & \cite{In_023_IRS_CO_2023, In_024_Hanzo_2020} \\ \cline{2-4}
 & Generative model  & It flexibly captures complex data distributions and handles non-convex problems, but requires an optimal labeled dataset. &  ~\cite{In_001_GAI_2025, In_002_DMbeatGAN ,In_003_GAI_2025, In_004_DM_2025, In_005_GAI_2024, In_006_DiffSG_2025,  In_011_Cond_DM_2025, In_012_CDDPM_letafati_2025, In_013_Cond_CDDM_2025, In_025_GCDM_2026}  \\
\cline{2-4}
 & Deep reinforcement learning (DRL)  & It learns without any labeled dataset, but it is inefficient in complex, high-dimensional environments due to extensive exploration. &  ~\cite{In_026_Capacity_DDQN_2023, In_027_DRL_RIS_2023, In_028_DRL_2023, In_029_Enhancing_2024} \\ \cline{2-4} 
  & Self-supervised fitness- driven diffusion (SFiD)  &  It does not require any labeled dataset and flexibly captures complex data distributions and it handles non-convex problems. & \textbf{Proposed}\\ 
\hline 
\end{tabular}
\vspace{-10 pt}
\end{table*}

Meta-heuristic methods such as genetic algorithm (GA) \cite{In_016_Rate-Splitting_2025, IN_017_GA_Power_2022} and particle swarm optimization (PSO)~\cite{In_018_PSO_Malay_2025} are capable of producing near-optimal or optimal solutions, but their computational burden and slow convergence may limit their practicality in real-world systems having rapidly time-variant conditions. Moreover, conventional optimization methods typically employ iterative procedures or rely on approximate convex reformulations to obtain feasible solutions. For instance, block coordinate descent (BCD)~\cite{In_019_EE_BCD_SCA_2023, In_020_BCD_2021, In_021_SCA_Phase_2025} and successive convex approximation (SCA)~\cite{In_019_EE_BCD_SCA_2023, In_020_BCD_2021, In_021_SCA_Phase_2025, In_022_SCA_2021} are iterative optimization methods that provide suboptimal solutions and involve complex mathematical derivations. BCD updates one coordinate or block at a time, often resulting in slow convergence in high-dimensional spaces and exhibits sensitivity to coordinate selection and step sizes. SCA addresses non-convex problems through sequential convex approximations, but requires careful formulation and convergence analysis, making its implementation challenging. Consequently, both methods are inefficient or unreliable for large-scale, high-dimensional or non-convex problems.

Deep learning based discriminative models~\cite{In_023_IRS_CO_2023, In_024_Hanzo_2020} offer a computationally efficient alternative to conventional optimization methods. However, they require large labeled datasets and rely on fixed input-output mappings. As a result, they often struggle in dynamic or highly non-convex environments, particularly when the test data distribution deviates from the training distribution~\cite{In_001_GAI_2025, In_004_DM_2025, In_006_DiffSG_2025}.
Consequently, deep reinforcement learning (DRL)~\cite{In_026_Capacity_DDQN_2023, In_027_DRL_RIS_2023, In_028_DRL_2023, In_029_Enhancing_2024}, which explores the solution space through interactions with the environment to make decisions, has gained increasing popularity, eliminating the need for labeled training data. However, the performance of reinforcement learning strongly depends on extensive interaction and exploration, which makes training difficult in high-dimensional and complex solution spaces and increases the likelihood of convergence to local optima~\cite{In_001_GAI_2025, In_004_DM_2025}. Alternatively, GAI~\cite{In_001_GAI_2025, In_002_DMbeatGAN, In_003_GAI_2025, In_004_DM_2025, In_005_GAI_2024, In_006_DiffSG_2025,  In_011_Cond_DM_2025, In_012_CDDPM_letafati_2025, In_013_Cond_CDDM_2025, In_025_GCDM_2026}, and in particular diffusion models, overcome these challenges by innovative data synthesis, modeling a wide range of realistic network conditions, and producing novel solutions in real time, making it eminently suitable for optimizing NG networks. In~\cite{In_029_Enhancing_2024}, Du et al. demonstrate that diffusion models trained on expert datasets achieve favorable performance and they recommend intrinsically integrating diffusion models with DRL to eliminate the need for external expert data. However, this integration leads to some performance erosion. Diffusion model based optimization achieves promising performance but relies on the availability of training datasets, which are not always accessible or may be difficult to generate. To overcome these challenges, we propose a self-learning diffusion model that eliminates the need for an explicit training dataset.


Observe from Table~\ref{Table-1} that  existing optimization approaches often rely on strong modeling assumptions, large labeled datasets, or problem-specific designs, which limit their adaptability and may result in suboptimal performance under rapidly fluctuating network conditions. By contrast, generative diffusion models exhibit strong generalization capability and maintain good performance, even when harnessed for data distributions that differ from those seen during training~\cite{In_006_DiffSG_2025}. Overall, the above methods either lack sufficient generalization capability or require high-quality training datasets.
Hence, we propose a novel framework that employs a diffusion-based optimization model, in which the  training data required is generated intrinsically
by the model itself, without relying on any external dataset generation mechanism. \textit{To the best of our knowledge, this is a novel technique of  tackling complex optimization problems while accounting for training data scarcity, eliminating the reliance on external training datasets, and allowing the model to self-generate the required data.} The proposed framework is applicable to a broad class of optimization problems. In this paper, we first validate its efficiency using widely adopted analytical benchmark functions having known global optima, which facilitates a clear and fair evaluation of optimization performance. We next investigate reconfigurable intelligent surface (RIS)-assisted cell-free massive multiple-input multiple-output (CF-mMIMO) systems under imperfect channel state information (CSI) and spatial correlation at the RIS, with the aim of RIS phase-shift optimization.
In NG wireless systems, RIS enables programmable amelioration of wireless propagation through adjustable reflecting elements, while CF-mMIMO employs a group of distributed access points (APs) for jointly serving a group of UEs without cell boundaries. Their integration provides significant gains in coverage, spectral efficiency and energy efficiency. The RIS phase-shift optimization problem is inherently non-convex and challenging. More explicitly, it involves either ideal continuous or practical discrete phase-shift variables, making the problem eminently well-suited for demonstrating the versatility of the proposed technique across different optimization domains. Against the above backdrop, the \textbf{key contributions} of this paper are as follows:

\begin{itemize}
\item We propose a novel technique, termed as self-supervised fitness-driven diffusion (SFiD), which generates its own training dataset for self-training without external data sources. We formulate a training objective that directs the model to converge toward optimal solutions by incorporating an optimization fitness function. The model is trained using data generated by itself, guided by the fitness function, thereby orchestrating a self-supervised learning process without relying on external datasets. This method allows the model to progressively refine its performance and efficiently explore the solution space.
\item We design the proposed framework to be highly generalizable, so that it remains compatible with a broad class of diffusion models and optimization tasks. We harness the denoising diffusion probabilistic model (DDPM)~\cite{In_009_DDPM} and propose SFiD-C for continuous-valued optimization problems. By contrast, for discrete-valued optimization, we adopt the binary diffusion probabilistic model (BDPM)~\cite{Mot_001_BDPM_2025} and develop SFiD-D. This is then further enhanced by incorporating a binary neural network (BNN)~\cite{Mot_002_BNN_survey_2023}, leading to our SFiD-D (BNN) scheme. 
 BNNs substantially reduce storage requirements, computational complexity, and energy consumption, which makes them highly amenable to compact and low-power hardware platforms.
 \item We first evaluate our proposed method on analytical mathematical optimization problems and then extend it to RIS phase-shift design in CF-mMIMO systems for maximizing the sum spectral efficiency (sum-SE). RIS phase optimization represents a challenging non-convex bottleneck in such systems, as it involves both continuous and discrete decision variables, making it an ideal test-case for demonstrating the flexibility and robustness of our technique.
\item {We validate the efficiency of our proposed method through extensive simulations. We evaluate our proposed methods, namely SFiD-C and SFiD-D, for RIS phase-shift optimization in CF-mMIMO systems against heuristic and DRL methods, demonstrating that they achieve comparable or superior performance. Our methods are capable of reducing the computational complexity of both continuous and discrete RIS phase-shift optimization by up to two orders of magnitude compared to pertinent baseline methods.}
\end{itemize}

\textit{{Notation:}} 
{Scalars are denoted by lower or uppercase italic letters (e,g. $a$, $A$), vectors by bold lowercase letters (e,g. $\mathbf{a}$, $\boldsymbol{\theta}$), matrices by bold uppercase letters (e,g. $\mathbf{A}$, $\boldsymbol{\Theta}$) and sets by calligraphic letters ($\mathcal{A}$). Furthermore,
$\mathbb{E}[\cdot]$ denotes the expectation operator, $(\cdot)^*$, $(\cdot)^T$ and $(\cdot)^H$ represent the conjugate, transpose and Hermitian transpose, respectively.
$\|\cdot\|$ denotes the Euclidean norm, $\oplus$ represents the XOR operation, $\textbf{I}_N$ denotes the $N \times N$ identity matrix,
$\ln(\cdot)$ and $\log_2(\cdot)$ denote the natural logarithm and base-2 logarithm, respectively. $ D_{\mathrm{KL}}(P \| Q) \triangleq \mathbb{E}_{x \sim P}
\left[ \ln \frac{P(x)}{Q(x)} \right]$ denotes the Kullback--Leibler (KL)
divergence between distributions $P$ and $Q$.
$\mathcal{N}(\boldsymbol{\mu}, \boldsymbol{\Sigma})$ and 
$\mathcal{CN}(\boldsymbol{\mu}, \boldsymbol{\Sigma})$ denote a Gaussian
distribution and a circularly symmetric complex Gaussian distribution,
respectively, with mean $\boldsymbol{\mu}$ and covariance $\boldsymbol{\Sigma}$.
}



\section{Self-Supervised Fitness-Driven Diffusion}

In this section, we introduce SFiD, a novel technique that enables diffusion-based models to self-optimize without external datasets by leveraging fitness-guided feedback mechanisms. We propose SFiD-C for continuous variable optimization, and SFiD-D for discrete variable optimization.
\vspace{-10pt}
\subsection{Concept and Mechanism}

Diffusion models have demonstrated compelling benefits in solving complex optimization tasks~\cite{In_004_DM_2025, In_006_DiffSG_2025}. However, their performance typically depends on large expert optimal datasets, which are often unavailable or costly to obtain in practical optimization scenarios. To circumvent this limitation, we utilize the optimization objective itself as the self-supervision guidance. This gives birth to a self-supervised fitness-driven diffusion mechanism in which the model iteratively samples, evaluates, and refines candidate solutions. This, in turn, steers the diffusion process toward optimal solutions without requiring an external expert training dataset.

Let us consider a generic optimization problem denoted by the fitness function $\mathcal{R}(\boldsymbol{\theta}, \mathcal{C})$, with $\boldsymbol{\theta}$ representing the optimization variable and  $\mathcal{C}$ denoting the set of conditional information and/or fixed parameters. Our objective is to maximize this fitness function under the given constraints, which can be generally expressed as
\begin{subequations}\label{GProblem}
 \begin{align} \quad
     \underset{\boldsymbol{\theta}}{\text{maximize}} \;\;& \mathcal{R}(\boldsymbol{\theta}, \mathcal{C}),\label{P1_a} \\
     \text{subject to} \;\; 
     &f_i(\boldsymbol{\theta}, \mathcal{C})\leq 0,  \;\; \forall i \in \mathcal{F},  \\
    &g_j(\boldsymbol{\theta}, \mathcal{C}) = 0, \;\; \forall j \in \mathcal{G},\\
    &h_k(\boldsymbol{\theta}, \mathcal{C}) \in \mathcal{D}, \;\; \forall k \in \mathcal{H},\label{P1_d}   
 \end{align} 
 \end{subequations}
 where $\mathcal{F}$, $\mathcal{G}$ and $\mathcal{H}$ are the index sets of
 the inequality functions, equality functions and discrete-set constraint functions, respectively. In SFiD, our goal is to train the model such that, for a given conditioning information $\mathcal{C}$, it generates an optimized $\boldsymbol{\theta}$ that maximizes the fitness value 
$\mathcal{R}(\boldsymbol{\theta}, \mathcal{C})$. In the following section, we derive the fitness driven objective that enables the model to learn directly from the optimization objective itself, eliminating the need for any expert dataset.
 
\vspace{-10pt}

\subsection{Fitness Driven Objective}
Our objective is to maximize 
$\mathcal{R}(\boldsymbol{\theta}, \mathcal{C})$ given the conditional information 
 $\mathcal{C}$. We aim for learning the conditional distribution $p(\boldsymbol{\theta} \! \mid \! \mathcal{C})$ so that, for any given 
$\mathcal{C}$, samples $\boldsymbol{\theta}$ drawn from this distribution yield maximal values of the fitness function $\mathcal{R}(\boldsymbol{\theta}, \mathcal{C})$. So, the objective function is formulated as 
\begin{equation}
   J = \max_{p_w(\boldsymbol{\theta} \mid \mathcal{C})} \ \mathbb{E}_{\mathcal{C},\boldsymbol{\theta} \sim p_w(\boldsymbol{\theta} \mid \mathcal{C})}[\mathcal{R}(\boldsymbol{\theta}, \mathcal{C})]. \label{eq:J} 
\end{equation}
We add an entropy regularization term to the objective function in~\eqref{eq:J} to prevent mode-collapse by maintaining diversity in the generated solutions. Briefly, mode-collapse refers to the tendency of the model to concentrate on a limited set of high-probability solutions, which limits exploration and leads to suboptimal solutions. The revised objective function is expressed as
\begin{equation}
   \tilde{J} = \max_{p_w(\boldsymbol{\theta} \mid \mathcal{C})}  \big\{\mathbb{E}_{\mathcal{C},\boldsymbol{\theta} \sim p_w(\boldsymbol{\theta} \mid \mathcal{C})}\big(\mathcal{R}(\boldsymbol{\theta}, \mathcal{C})\big) + \gamma \mathbb{E}_{\mathcal{C}}[H\big(p_w(\boldsymbol{\theta} \mid \mathcal{C})\big)] \big\} \label{eq:J_tilda}
\end{equation}
where, $ H(p_w(\boldsymbol{\theta} \mid \mathcal{C})) \triangleq -\mathbb{E}_{\boldsymbol{\theta} \sim p_w(\boldsymbol{\theta} \mid \mathcal{C})}[\ln{p_w(\boldsymbol{\theta} \mid \mathcal{C})}] $ denotes the entropy  (uncertainty in generating $\boldsymbol{\theta}$, given the conditional information $\mathcal{C}$). The parameter $\gamma >0$ controls the trade-off between exploitation and exploration: a small $\gamma$ encourages aggressive exploitation by concentrating the probability mass on high-fitness values, while a large $\gamma$ promotes exploration by keeping the distribution broader. This secondary term encourages diversity in the generated solutions by penalizing the model, when it concentrates excessively on a limited set of modes. Consequently,~\eqref{eq:J_tilda} studies a trade-off between optimizing for fitness and maintaining a rich representation of diverse solutions, thereby improving the robustness and generalization of the learned distribution's capability.
\begin{theorem} \label{Theorem-1}
The optimal solution to the fitness maximization objective in \eqref{eq:J_tilda} is given by
\begin{equation}
     p^*_w(\boldsymbol{\theta} \mid \mathcal{C}) \propto \exp{\left(\frac{1}{\gamma} \mathcal{R}(\boldsymbol{\theta}, \mathcal{C})\right)}.
\end{equation}

\end{theorem}
\begin{proof}
The objective function in \eqref{eq:J_tilda} is simplified as follows
   \begin{align}
   \tilde{J} &= \max_{p_w(\boldsymbol{\theta} \mid \mathcal{C})} \ \mathbb{E}_{\mathcal{C}} \left[
   \mathbb{E}_{\boldsymbol{\theta} \sim p_w(\boldsymbol{\theta} \mid \mathcal{C})} \big( \mathcal{R}(\boldsymbol{\theta}, \mathcal{C}) \big) 
   + \gamma \, H\!\big(p_w(\boldsymbol{\theta} \mid \mathcal{C})\big) \right], \notag\\
   &= \min_{p_w(\boldsymbol{\theta} \mid \mathcal{C})} \ \mathbb{E}_{\mathcal{C}} \left[ 
   \mathbb{E}_{\boldsymbol{\theta} \sim p_w(\boldsymbol{\theta} \mid \mathcal{C})} \left(\ln\frac{p_w(\boldsymbol{\theta} \mid \mathcal{C})}{\exp{\big(\frac{1}{\gamma} \mathcal{R}(\boldsymbol{\theta}, \mathcal{C})\big)} }\right) \right]. \label{eq:obj1}
\end{align}
Now we define $p^*_w(\boldsymbol{\theta} \ | \ \mathcal{C}) \triangleq  \frac{1}{Z(\mathcal{C})} \exp{\big(\frac{1}{\gamma} \mathcal{R}(\boldsymbol{\theta}, \mathcal{C})\big)}$,~where $Z(\mathcal{C})$ is a normalizing factor, formulated as $Z(\mathcal{C})=\int_{\boldsymbol{\theta}} \exp{\big[\frac{1}{\gamma} \mathcal{R}(\boldsymbol{\theta}, \mathcal{C})\big] } \, d\boldsymbol{\theta} $ and $Z(\mathcal{C}) = \sum_{\boldsymbol{\theta}} \exp{\big[\frac{1}{\gamma} \mathcal{R}(\boldsymbol{\theta}, \mathcal{C})\big]}$ for continuous and discrete optimization variables, respectively. Accordingly, \eqref{eq:obj1} is written as
\begin{align}
   \tilde{J} &= \min_{p_w(\boldsymbol{\theta} \mid \mathcal{C})} \ \mathbb{E}_{\mathcal{C}}\left[  
   \mathbb{E}_{\boldsymbol{\theta} \sim p_w(\boldsymbol{\theta} \mid \mathcal{C})} \left(  \ln\frac{p_w(\boldsymbol{\theta} \mid \mathcal{C})}{p^*_w(\boldsymbol{\theta} \mid \mathcal{C}) } - \ln{Z(\mathcal{C})} \right)\right], \notag\\
    &=\min_{p_w(\boldsymbol{\theta} \mid \mathcal{C})}\left[ \mathbb{E}_{\mathcal{C}} \left[ D_{\mathrm{KL}}(p_w \, \| \, p^*_w)\right] -  \mathbb{E}_{\mathcal{C},\boldsymbol{\theta} \sim p_w(\boldsymbol{\theta} \mid \mathcal{C})}\left[\ln Z(\mathcal{C}) \right] \right].
\end{align}
Since $Z(\mathcal{C})$ does not depend on  $p_w(\boldsymbol{\theta}\!\mid\!\mathcal{C})$, the minimization reduces to
\begin{align}
   \bar{J} 
    &= \min_{p_w(\boldsymbol{\theta} \mid \mathcal{C})} \  \mathbb{E}_{\mathcal{C}} \left[ D_{\mathrm{KL}}(p_w \, \| \, p^*_w)\right].
\end{align}
From Gibbs’ inequality~\cite{Gibb_BremAud2012}, the KL divergence achieves its minimum value of zero if and only if the two probability distributions are identical. Therefore, the optimal solution is given by
\begin{align}
    p_w(\boldsymbol{\theta} \mid \mathcal{C}) &= p^*_w(\boldsymbol{\theta} \mid \mathcal{C}) = \frac{1}{Z(\mathcal{C})} \exp{\left(\frac{1}{\gamma} \mathcal{R}(\boldsymbol{\theta}, \mathcal{C})\right)}. \notag 
\end{align}
Hence, the optimal solution obeys
\begin{align}
    p^*_w(\boldsymbol{\theta} \mid \mathcal{C}) &\propto  \exp{\left(\frac{1}{\gamma} \mathcal{R}(\boldsymbol{\theta}, \mathcal{C})\right)}. \label{eq:FOD_final}
\end{align}
\end{proof}

It follows from \eqref{eq:FOD_final} that solutions having higher fitness are exponentially more likely to be selected. Thus, sampling from this distribution naturally biases the model toward high fitness regions of the parameter space, enabling optimization through probabilistic generation, rather than explicit search. In the following subsection, we derive the normalized weight formulation that enables fitness driven diffusion process during SFiD model training.

\vspace{-10pt}

\subsection{Normalized Weight Formulation for SFiD Training}
\begin{figure}[htbp]
\includegraphics[width=1\linewidth,height=0.5\linewidth]{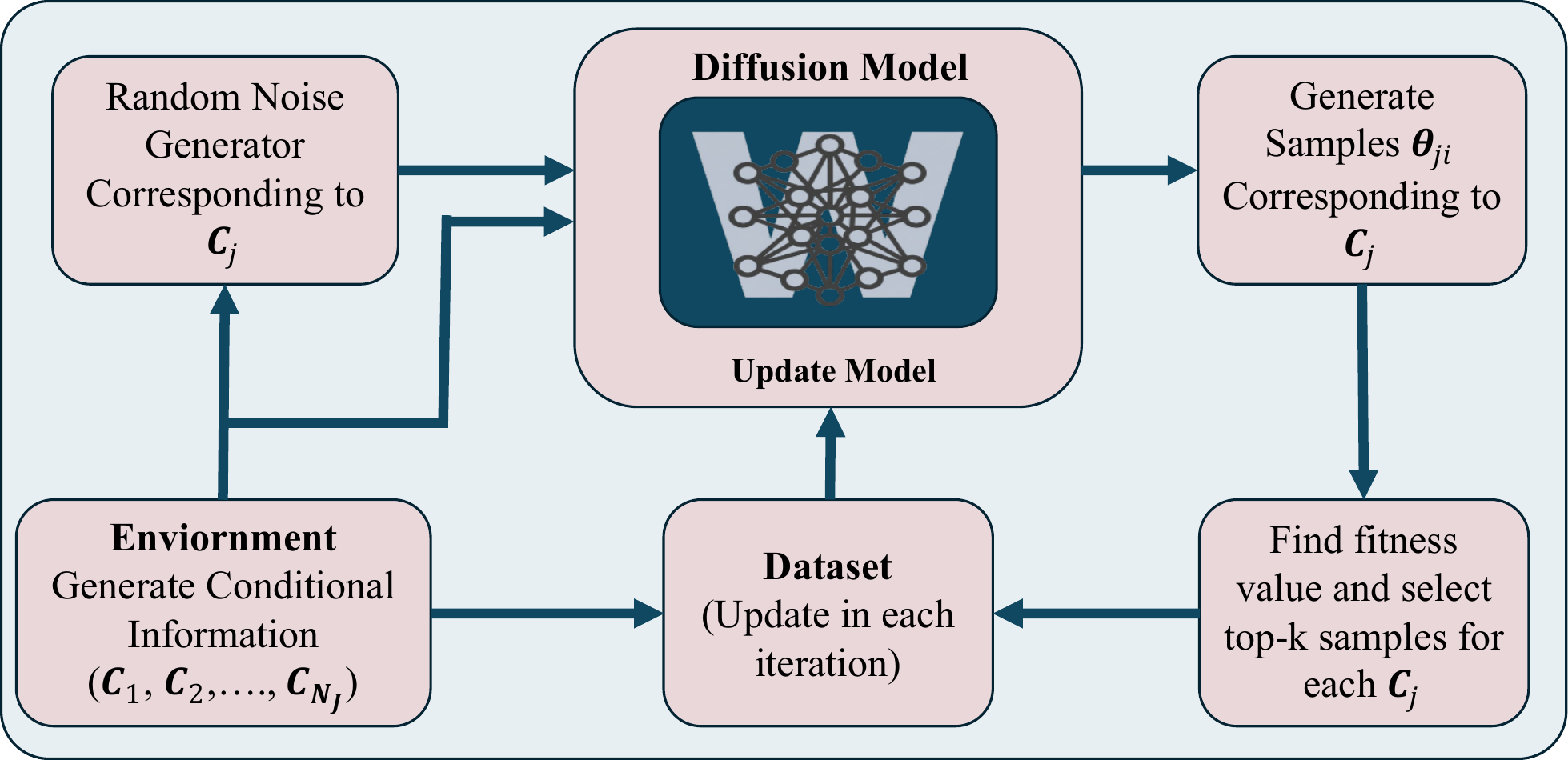} 
\vspace{-15pt}
\caption{\small The proposed SFiD framework.} 
 \label{Fig1:SFiD model}
\end{figure}

We train our model by guiding it toward the optimal distribution derived in \eqref{eq:FOD_final}. This distribution assigns a higher probability to samples associated with higher fitness values, where the fitness function 
$\mathcal{R}(\boldsymbol{\theta}, \mathcal{C})$ shapes the search landscape and the parameter 
$\gamma$ controls the balance between exploration and exploitation. Here, we generate samples from the model, evaluate their fitness values, and assign higher importance to samples that perform better. These weighted samples are then used for fine-tuning the model, gradually steering it towards generating more promising candidates, while still allowing for population diversity and exploration, as shown in Fig.~\ref{Fig1:SFiD model}.

First, we generate the samples $\boldsymbol{\theta}_{ji}$ for a given condition $\mathcal{C}_j$, where $j=1,2,\dots,N_J$ and $i=1,2,\dots\!,N_I$ with $N_J$ denoting the number of conditional information instances and 
$N_I$ representing the number of samples generated per conditional information. Next, we compute the fitness value for each $\boldsymbol{\theta}_{ji}$ as $r_{ji} = \mathcal{R}(\boldsymbol{\theta}_{ji}, \mathcal{C}_j)$, and then determine the corresponding weight $\lambda_{ji}$ using \eqref{eq:FOD_final}, given by
\begin{equation}
    \lambda_{ji} = \exp{\left(\frac{1}{\gamma} \mathcal{R}(\boldsymbol{\theta}_{ji}, \mathcal{C}_j)\right)} = \exp{\left(\frac{r_{ji}}{\gamma} \right)}. 
\end{equation}
Now, we normalize the weights and, instead of reweighting all samples, we retain only the $\text{top-}k$ samples of the highest fitness values, focusing training on the most promising candidates while ignoring the rest. Here, normalized weights are used for ensuring numerical stability and a consistent probabilistic interpretation of the fitness values. Furthermore, restricting the optimization to the top-$k$ candidates focuses learning on high-quality solutions reduces the impact of low-fitness samples, and improves convergence efficiency. Thus, the normalized weight is defined as
\begin{equation}
\label{eq:normalized_weight}
\tilde{\lambda}_{ji} = 
\begin{cases}
\dfrac{{\lambda}_{ji}}{\sum\limits_{i \in \text{top-}k} {\lambda}_{ji}}, & \text{if } i \in \text{top-}k, \\[10pt]
\ \ 0, & \text{otherwise}.
\end{cases}
\end{equation}

Based on this SFiD framework, a pair of specialized formulations is developed for different optimization domains. Explicitly, SFiD-C and SFiD-D are designed for continuous-variable and discrete-variable optimization, respectively. Both variants retain the core self-supervised, fitness-driven learning mechanism of SFiD, while harnessing the diffusion and sampling processes for effectively handling the unique characteristics of continuous- and discrete-valued optimization variables.

\vspace{-10pt}
\subsection{SFiD-C: Continuous-Variable Optimization}
In this section, we present SFiD-C, which addresses continuous-variable optimization based on continuous diffusion models~\cite{In_009_DDPM,  In_011_Cond_DM_2025, In_012_CDDPM_letafati_2025}. Diffusion models directly learn the underlying data distribution $p(\boldsymbol{\theta})$ from samples $\boldsymbol{\theta}_0$ and subsequently generate new samples through the sampling process. 
Moreover, conditional diffusion models learn 
$p(\boldsymbol{\theta} \! \mid \! \mathcal{C})$ based on a given condition $\mathcal{C}$. The diffusion process gradually adds Gaussian noise to a sample $\boldsymbol{\theta}_0$, producing $\boldsymbol{\theta}_t$ over time steps $t \in \{1,2,\dots,T\}$ until the final distribution $p_T(\boldsymbol{\theta}_{T})$ becomes pure noise. The forward transition at each step is modeled as 
\begin{equation}
p(\boldsymbol{\theta}_t|\boldsymbol{\theta}_{t-1}, \mathcal{C}) \sim \mathcal{N}(\boldsymbol{\theta}_t;  \sqrt{m_t} \boldsymbol{\theta}_{t-1}, v_t \mathbf{I}_N),
\label{eq:forward_process1}
\end{equation}
where $m_t$ and $v_t$ represent the hyperparameters of the predefined noise schedule~\cite{In_009_DDPM}. Through this process, the model captures the noise distribution, which allows it to progressively denoise an initial distribution toward the target distribution. For the denoising process, a neural network is employed for parameterizing the reverse distribution, formulated as $ p_w(\boldsymbol{\theta}_{0:T}) = \prod_{t=1}^T p_w(\boldsymbol{\theta}_{t-1} \mid \boldsymbol{\theta}_t),
$ where we have
\begin{equation}
p_w(\boldsymbol{\theta}_{t\!-\!1}|\boldsymbol{\theta}_{t}, \mathcal{C}) \sim \mathcal{N}(\boldsymbol{\theta}_{t-1}; \boldsymbol{\mu}_w(\boldsymbol{\theta}_{t},t, \mathcal{C}), \boldsymbol{\Sigma}_w(\boldsymbol{\theta}_{t}, t, \mathcal{C})).
\label{eq:Reverse_process_NN}
\end{equation}
The variational upper bound of the negative log-likelihood is adopted as the loss function for learning the original data distribution and it is then further simplified to the following loss function~\cite{In_009_DDPM}, in which the SFiD normalized weight $\tilde{\lambda}$ in~\eqref{eq:normalized_weight} is incorporated as
\begin{equation} \label{eq:DM_ELBO}
    \mathcal{L}_{\text{C}} = \mathbb{E}_{} 
    \left[ \tilde{\lambda} \cdot\left\| \boldsymbol{\epsilon} - \boldsymbol{\epsilon}_w(\boldsymbol{\theta}_{t}, t, \mathcal{C}) \right\|^2 \right],
\end{equation}
where $\boldsymbol{\epsilon} \sim \mathcal{N}(0,\mathbf{I}_N)$ represents Gaussian noise and $\boldsymbol{\epsilon}_w$ denotes the neural network's prediction. 

\vspace{-10pt}
\subsection{SFiD-D: Discrete-Variable Optimization}

SFiD-C, which is built upon Gaussian noise and continuous latent representations, is inherently limited when the optimization variables are discrete or binary, making it unsuitable for directly modeling such discrete search spaces. In SFiD-C, Gaussian noise operates under the assumption of a continuous data space, making it well-suited for real-valued data, but rendering it ineffective for binary or discrete inputs~\cite{Mot_001_BDPM_2025}. When applied to binary data, Gaussian noise produces intermediate continuous values that must be quantized or directly binarized to preserve the binary format. This process often introduces artifacts, causes information loss, and hinders convergence, revealing a fundamental mismatch between Gaussian noise and the inherently discrete nature of digital data. BDPM~\cite{Mot_001_BDPM_2025} are specifically designed for discrete or binary data and maintain the binary nature of the data throughout the entire diffusion process, hence eliminating the need for quantization or binarization. This approach ensures bit-level consistency, prevents information loss, and provides a more natural fit for binary data compared to Gaussian-based diffusion models. Therefore, we propose SFiD-D, which leverages a BDPM for effectively handling binary/discrete optimization variables.

In SFiD-D, the perturbation is injected using the XOR operation, which makes the model naturally compatible with binary data. This mechanism ensures that the perturbed samples remain within the binary domain throughout the entire diffusion process, preserving the discrete structure of the data. The perturbation $\boldsymbol{z}^b_t$ is introduced by flipping bits through the XOR operation, where the flip probability is defined as a function of the timestep $t$ and lies within the range $[0, 0.5]$. Let $\boldsymbol{\theta}^b_0$ denote the original binary data. The perturbed binary data at timestep $t$ is then given by
\begin{equation} \label{eq:BDPM_noise_added}
    \boldsymbol{\theta}^b_t = \boldsymbol{\theta}^b_0 \oplus \boldsymbol{z}^b_t.
\end{equation}

The denoising neural network predicts the added perturbation $\boldsymbol{z}^b_t$ as $\boldsymbol{z}^b_w(\boldsymbol{\theta}^b_{t}, t, \mathcal{C})$, by exploiting the perturbed binary sample $\boldsymbol{\theta}^b_{t}$, the timestep $t$, and the conditional information $\mathcal{C}$ as its inputs. Explicitly, the model learns by minimizing the loss function
\begin{equation} \label{eq:BM_loss}
    \mathcal{L}_{\text{B}} = \mathbb{E}_{} 
    \left[ \tilde{\lambda}  \cdot\left\| \boldsymbol{z}^b_t -\boldsymbol{z}^b_w(\boldsymbol{\theta}^b_{t}, t, \mathcal{C})\right\|^2 \right],
\end{equation}
where, $\tilde{\lambda}$ is the normalizing weighting factor defined in \eqref{eq:normalized_weight}.

In this work, we enhance SFiD-D by incorporating a BNN for improving its efficiency, which we refer to as SFiD-D (BNN). In SFiD-D, a BNN is employed as the denoising model, leveraging recent advances that have demonstrated the efficiency of BNNs in resource-constrained environments by significantly reducing storage requirements, computational cost, and energy consumption, making them ideal for compact and low-power devices~\cite{Mot_002_BNN_survey_2023}.  We employ $1$-bit activations and $1$-bit weights in the SFiD-D (BNN) to ensure efficient computation and maintain the binary nature of the diffusion process. The binarization process reduces the model’s storage requirements and additionally, it reduces the computational cost by replacing standard matrix multiplications with efficient XNOR and popcount operations. SFiD-D (BNN) can achieve up to $32\times$ lower memory usage and $58\times$ faster convolution operations compared to SFiD-D \cite{Mot_002_BNN_survey_2023, BNN_new_2024}. The SFiD-D (BNN) introduces binarization steps for both activations ${\theta}_{a}$ and weights $w$ in its forward path prior to performing the convolution operation. The Sign function is commonly employed to perform the binarization process and defined as
\begin{equation} \label{eq:sign_fun}
    \operatorname{Sign}(x) =
\begin{cases}
+1, & \text{if } x \ge 0,\\
-1, & \text{otherwise.}
\end{cases}
\end{equation}
After binarization, the activations 
${\theta}_{a}$ and weights 
$w$ are expressed as
\begin{equation} \label{eq:binarization}
{\theta}^b_{a} = \operatorname{Sign}({\theta}_{a}) \quad \text{and} \quad
w^b = \operatorname{Sign}(w).
\end{equation}
Since the derivative of the binarization function (Sign) is zero, binary weights cannot be optimized using the traditional gradient descent method of back-propagation. To overcome this limitation, we use the popular clipped straight-through estimator (STE) technique~\cite{Mot_002_BNN_survey_2023}, allowing gradient information to pass through the non-differentiable binarization function during the backward propagation process. The clipped STE function is defined~as 
\begin{equation} \label{eq:Clipped_STE}
\mathbb{I}_{|x| \le 1} =
\begin{cases}
1, & \text{if } -1 \leq x \leq 1,\\
0, & \text{otherwise.}
\end{cases}
\end{equation}
Although SFiD-D (BNN) provides prompt inference and a smaller model size, making them well-suited for resource-constrained devices, their training still relies on conventional high-performance hardware and incurs computational costs and training times comparable to SFiD-D. The next sub-section provides an illustrative example of how the SFiD framework approaches the optimal solution.

\vspace{-10pt}
\subsection{Conceptual Illustration of SFiD} 
\vspace*{-3pt}

\begin{figure}[htbp]
\includegraphics[width=0.99\linewidth,height=0.55\linewidth]{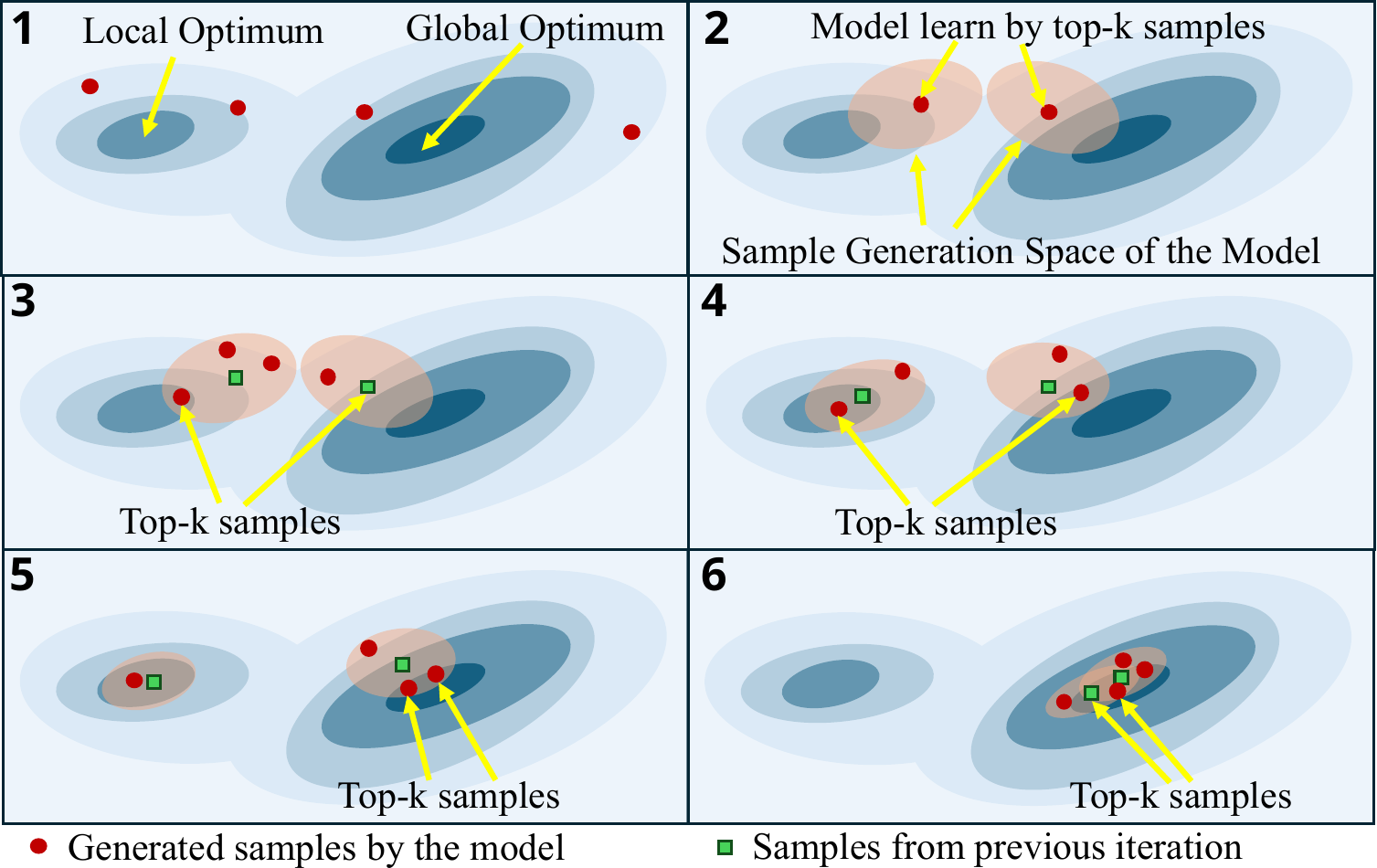}
\vspace{-20 pt}
\caption{\small Conceptual illustration of SFiD.} 
 \label{Fig2:Illustration of SFiD}
\end{figure}

Fig.~\ref{Fig2:Illustration of SFiD}  illustrates a conceptual characterization of the proposed SFiD method. In Step $1$, the solution landscape for the given condition is illustrated, where darker regions represent higher fitness values and lighter regions correspond to progressively lower fitness values. Here, the conditional information instances $N_J$ are set to $1$, since the illustration is provided for a single condition; hence, for notational simplicity, the subscript $j$ in $\boldsymbol{\theta}_{ji}$ and $\mathcal{C}_j$ is omitted. The model-generated samples $\boldsymbol{\theta}_{1}, \dots, \boldsymbol{\theta}_{4}$ are highlighted using red circles, {with the number of samples generated per conditional information $N_I$ set to $4$.} Next, we evaluate the fitness $r_{i} = \mathcal{R}(\boldsymbol{\theta}_{i}, \mathcal{C})$ for each $\boldsymbol{\theta}_{i}$, retain the top-$k$ performers ($k=2$), and compute the normalized weights as defined in~\eqref{eq:normalized_weight}. These weights are then used for learning the model, as shown in Step $2$. As shown in Step $3$, we generate new samples using the learned model, while retaining the top-$k$ samples inferred from previous steps, represented by green squares. These steps are then repeated, following the procedure illustrated in Steps $2$ and $3$. Note that from the second iteration onward, the top-$k$ samples for the current iteration are evaluated by considering both the retained top-$k$ samples from the previous iteration and the newly generated samples from the current iteration. Since the top-$k$ samples gleaned from the previous iteration are retained, each subsequent iteration maintains a set of high-fitness solutions whose fitness values are guaranteed to be equal to or higher than those in the preceding iteration. At the same time, the SFiD model generates new candidate solutions by sampling from the learned distribution, thereby enabling exploration of previously unvisited regions of the solution space. As iterations progress, this combination of exploitation and exploration increases the probability of identifying high-quality solutions and progressively approaching the optimum. This iterative refinement process is illustrated in Steps $4$-$6$ of Fig.~\ref{Fig2:Illustration of SFiD}, where the sample generation space of the model gradually shifts toward regions of higher fitness while still maintaining diversity to explore alternative optima.

\begin{figure*}[htbp]
	\centering
	\begin{subfigure}[b]{0.30\linewidth}		\hspace{-25pt}\includegraphics[width=1.2\linewidth,height=0.9\linewidth]{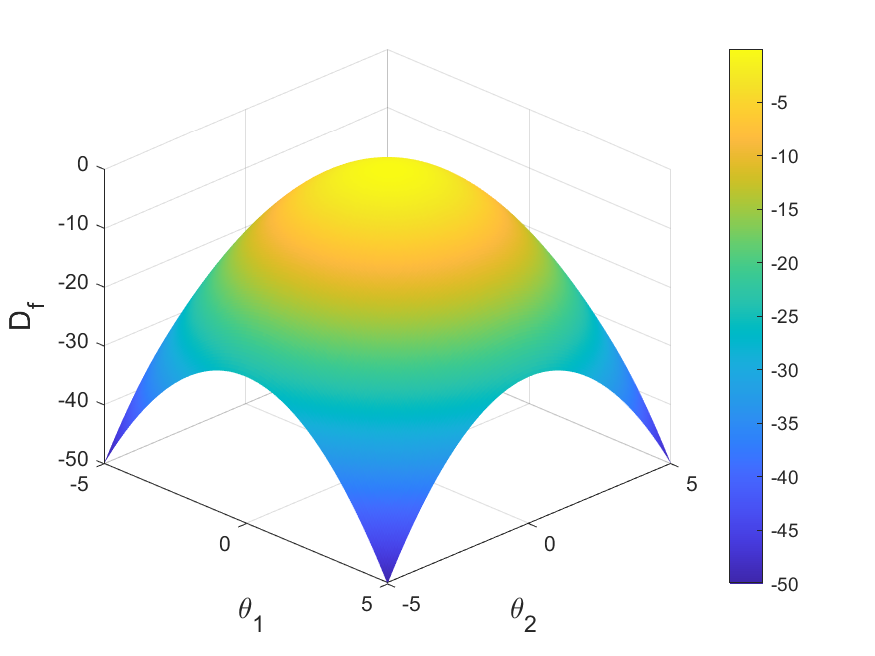} 
		\caption{\small } 
		\label{fig10a:DeJong_3D}
	\end{subfigure}\vspace{-2pt}
	\begin{subfigure}[b]{0.31\linewidth}
    
    \hspace{-12pt}\includegraphics[width=1.2\linewidth,height=0.9\linewidth]{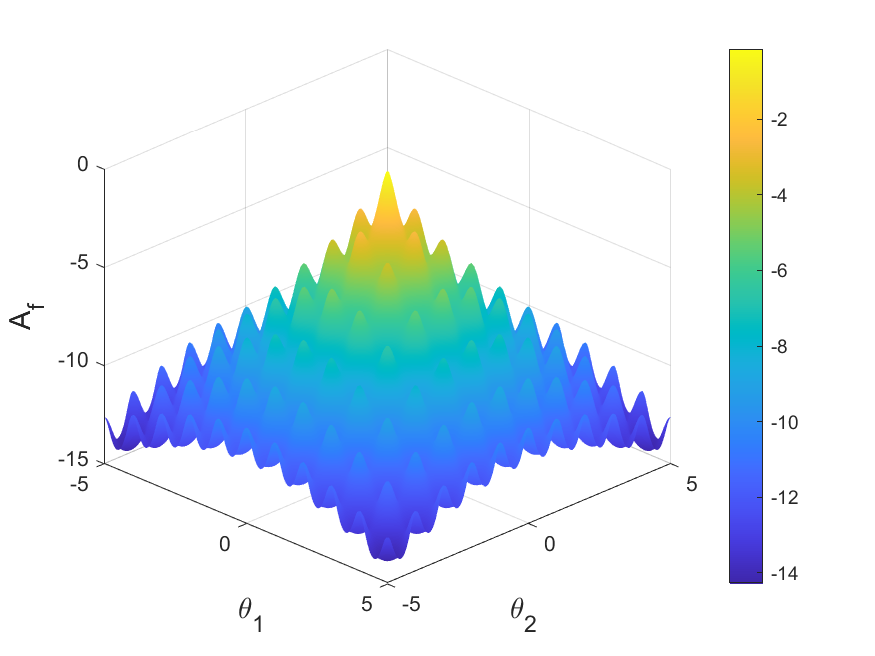}
	\caption{\small } 
	\label{fig10b:Ackley_3D}
	\end{subfigure}\vspace{-2pt}
	\begin{subfigure}[b]{0.30\linewidth}
	\hspace{2pt}\includegraphics[width=1.2\linewidth,height=0.9\linewidth]{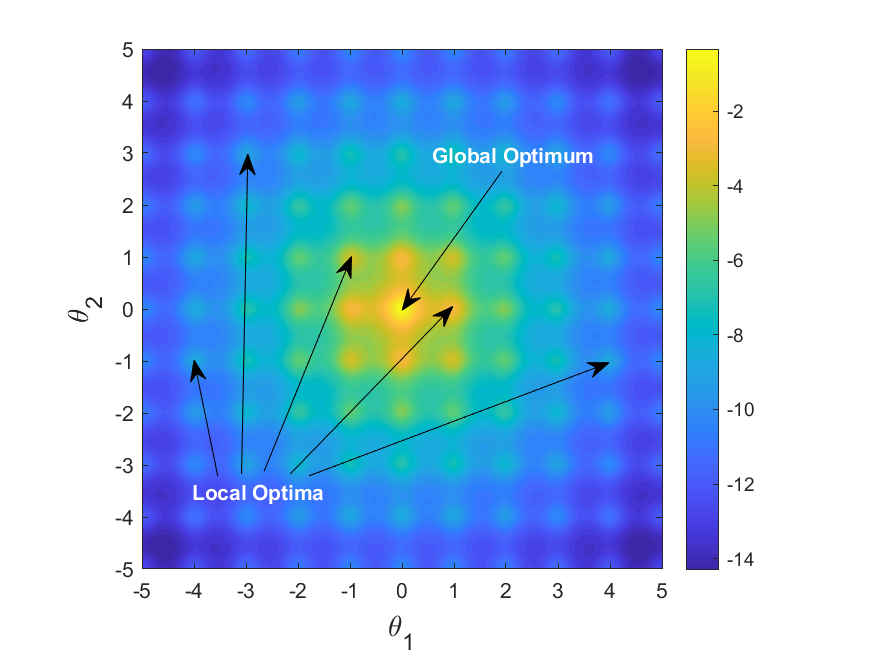}
	\caption{\small } 
	\label{fig10c:Ackley_2D}
	\end{subfigure} \vspace{-5pt}
	\caption{\small Visualization of the optimization function landscapes: a) De Jong's function, b) Ackley's function, c) 2D Contour of Ackley's function. 
    \label{fig10:Dejong_and_Ackley}\\[-20pt]} 
\end{figure*} 
\vspace{-15 pt}
\begin{figure*}[htbp]
	\centering
	\begin{subfigure}[b]{0.30\linewidth}		\hspace{-12pt}\includegraphics[width=1.1\linewidth,height=0.9\linewidth]{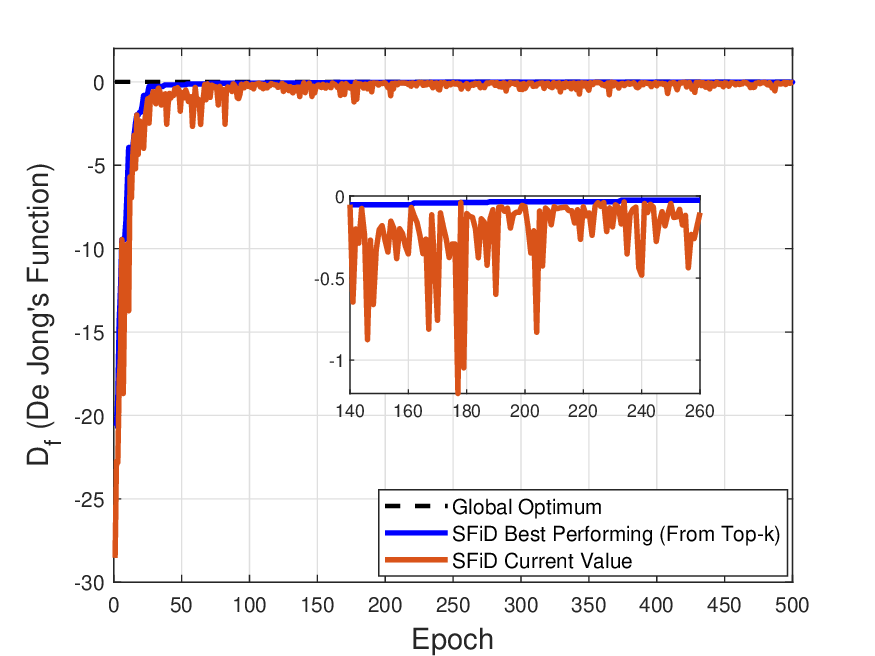} 
		\caption{\small} 
		\label{fig4a:Converg_DeJong}
	\end{subfigure}\vspace{-2pt}
	\begin{subfigure}[b]{0.30\linewidth}
    
    \hspace{-5pt}\includegraphics[width=1.1\linewidth,height=0.9\linewidth]{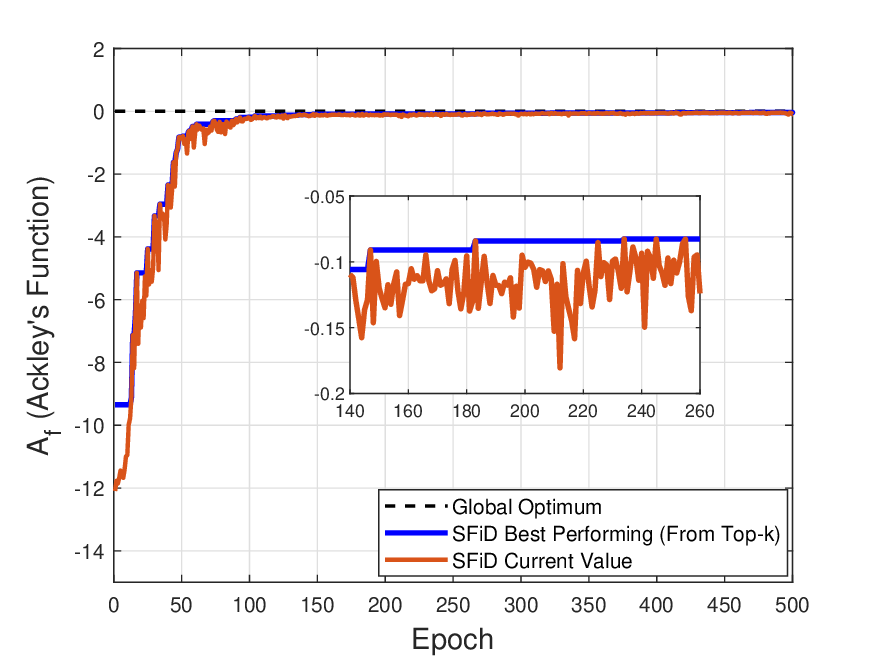}
	\caption{\small} 
	\label{fig4b:Converg_Ackley}
	\end{subfigure}\vspace{-2pt}
	\begin{subfigure}[b]{0.35\linewidth}
	\includegraphics[width=1.1\linewidth,height=0.78\linewidth]{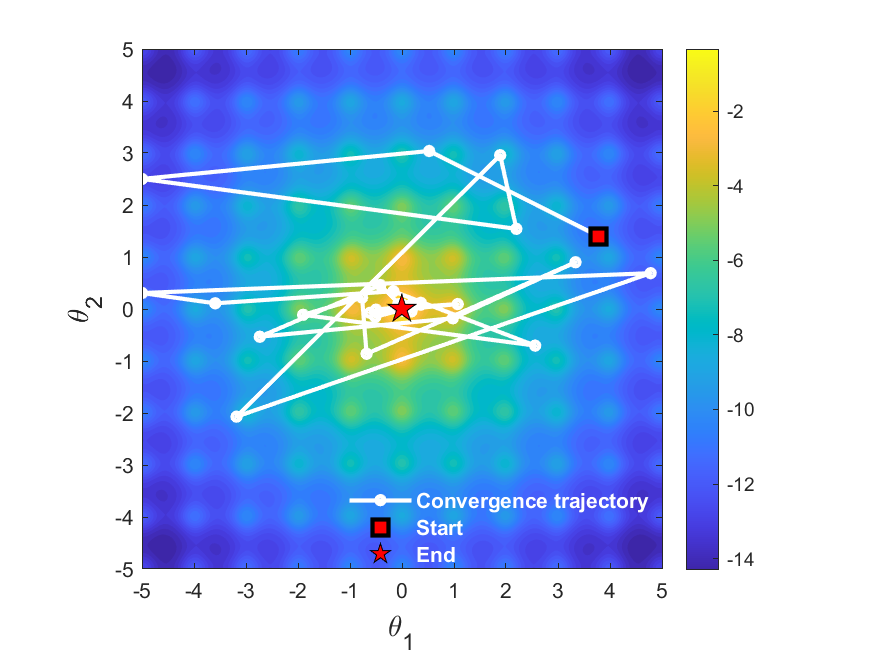}
	\caption{\small} 
	\label{fig4c:Converg_Ackley_2D}
	\end{subfigure} \vspace{-2pt}
	\caption{\small a) Convergence performance of our approach on De Jong’s functions, b) Convergence performance of our approach on Ackley’s functions, c) Convergence trajectory on the 2D contour of the Ackley's function.  
    \label{fig4:Converg_Dejong_and_Ackley}\\[-20pt]} 
\end{figure*}

\vspace{10 pt}
Normalizing the weights using the top-$k$ samples in \eqref{eq:normalized_weight} mitigates the impact of the parameter $\gamma$, particularly when the fitness values of the selected top-$k$ samples exhibit limited variation. At the early stages of training, the model generates samples having relatively high variance, which gradually decreases as the training progresses. Consequently, the influence of the parameter $\gamma$ is more pronounced during the initial training phase and diminishes as the model converges. To encourage sufficient exploration in the early stages, a relatively large value of $\gamma$ is therefore desirable. However, as the model learns and the sample variance reduces, the effective impact of $\gamma$ naturally decreases, reducing the need for further adjustment.  It is also worth noting that exploration is inherently handled by the model itself through probabilistic sample generation.    At the initial iterations, the model is still learning the solution distribution, so it generates highly diverse samples. But as the iterations progress, the model gradually shifts toward regions with higher fitness values, resulting in increased exploitation and reduced exploration. This exploration and exploitation enable the method to efficiently converge towards high-quality solutions.

Thus far, we have introduced the proposed SFiD framework in a general form, applicable to a wide class of optimization problems. In the next section, we first apply the proposed technique to widely used analytical benchmark functions, namely De Jong’s and Ackley’s~\cite{DeJong1}, with known optimal solutions, enabling a clear and fair assessment of optimization performance. Then, we focus our attention on a specific and practically important application, namely RIS phase-shift optimization in RIS-assisted CF-mMIMO systems. This problem has been well-studied in the literature, with both theoretical continuous phase-shift and practical discrete phase-shift implementations. Since RIS phase-shift optimization is inherently non-convex and challenging in both continuous and discrete domains, it serves as a test-case for evaluating the benefits of the proposed SFiD-C and SFiD-D algorithms.

\vspace{-10pt}

\section{Analytical Benchmark Functions} 
\vspace*{-3pt}
This section provides an overview of both the De Jong's and Ackley's benchmark functions and its convergence behaviour for our proposed approach. For both cases, the optimal solution is located at $\boldsymbol{\theta}=\boldsymbol{0} $. However, the corresponding loss landscapes are markedly different. De Jong’s function is characterized by a smooth landscape having a single local optimum, which constitutes a global optimum. By contrast, Ackley’s function exhibits a highly non-convex landscape with numerous local optima and a global optimum at $\boldsymbol{\theta}=\boldsymbol{0} $.  The De Jong’s function $(D_f)$ and the Ackley’s function $(A_f)$ for an $N$-dimensional variable $\boldsymbol{\theta}$ are mathematically formulated as follows~\cite{DeJong1}\vspace{-5 pt}
\begin{equation}
    D_f(\boldsymbol{\theta}) = (-1)\cdot\sum_{n=1}^{N} \theta_n^2 , \qquad \text{ and} \label{DeJong}  
\end{equation}
\begin{equation}
\begin{aligned}
A_f(\boldsymbol{\theta}) ={}& 20 \exp\!\left(
-0.2 \sqrt{\frac{1}{N} \sum_{n=1}^{N} \theta_n^2}
\right) \\
&\quad + \exp\!\left(
\frac{1}{N} \sum_{n=1}^{N} \cos(2\pi \theta_n)
\right)
- 20 - e, \label{Ackley}
\end{aligned}
\end{equation}
respectively. Note that both functions are originally formulated as minimization problems. In this work, the objective functions in~\eqref{DeJong} and~\eqref{Ackley} are multiplied by $-1$ to obtain equivalent maximization formulations. The optimization problem is defined as 
\begin{subequations}\label{Problem_P3&4}
\begin{align} \textbf{P1:}\quad
     \underset{\boldsymbol{\theta}}{\text{maximize}} \;\;& D_f(\boldsymbol{\theta}), \notag \\
     \text{subject to}\;\; 
     &\theta_{i} \in [-\zeta, \ \zeta], \;\;\; \forall i \in \{1, 2,\cdots, N\} \label{P3},       
 \end{align} 
  and
  \begin{align} \textbf{P2:}\quad
     \underset{\boldsymbol{\theta}}{\text{maximize}} \;\;& A_f(\boldsymbol{\theta}), \notag \\
     \text{subject to}\;\; 
     &\theta_{i} \in [-\zeta,\ \zeta], \;\;\; \forall i \in \{1, 2,\cdots, N\} \label{P4}.       
 \end{align} 
 \end{subequations}
Both functions are plotted in Fig.~\ref{fig10:Dejong_and_Ackley} for $N=2$ and $\zeta=5$. Again, Fig.~\ref{fig10a:DeJong_3D} shows that De Jong’s function has a single global optimum at $\boldsymbol{\theta}=\boldsymbol{0} $, whereas Fig.~\ref{fig10b:Ackley_3D} and Fig.~\ref{fig10c:Ackley_2D} indicate that Ackley’s function exhibits multiple local optima in addition to a global optimum at $\boldsymbol{\theta}=\boldsymbol{0} $.
\vspace{-5 pt}

\subsection{Convergence Analysis of De Jong's and Ackley's Functions}
{Fig.~\ref{fig4:Converg_Dejong_and_Ackley} illustrates the convergence behavior of the proposed SFiD-D technique for both De Jong's and Ackley's benchmark functions.
The results demonstrate that the proposed approach converges to the global optimum for the aforementioned benchmark functions. We perform the optimization using $N=64$ for both functions. Fig.~\ref{fig4a:Converg_DeJong} and Fig.~\ref{fig4b:Converg_Ackley} show the convergence curves for the De Jong's and Ackley's functions, respectively. The figures present three plots: (i) the global optimum value, (ii) the objective function value obtained from the optimization variables generated by the model at each iteration, shown as the SFiD current value (among the eight computed objective function values, each corresponding to a unique set of optimized variables generated by the model at the current iteration, only one representative value is presented for clarity), and (iii) the best-performing optimization variable set selected from the top-$k$ set. Both functions converge to near-optimal values within $200$ epochs. Due to its wider value range and smoother landscape, De Jong’s function requires more epochs to approach the vicinity of the optimum. However, once close enough, it reaches the actual optimal value faster than Ackley's function. By contrast, the highly rugged landscape of the Ackley's function exhibits slower convergence to the global optimum despite its smaller value range. For comparable maximum objective values, De Jong’s function converges in approximately $248$ epochs, whereas Ackley's function requires around $758$ epochs. This difference is justified by the highly rugged landscape of Ackley's function, which contains numerous local optima that slow down the convergence process. Additionally, Fig.~\ref{fig4c:Converg_Ackley_2D} demonstrates how the proposed approach converges on the 2D contour of Ackley's function for the first two dimensions. Overall, the proposed technique successfully converges for both functions, while requiring different training times depending on the shape of the objective function landscape.}

\section{{RIS phase-shift Optimization with SFiD}}
In this section, we carry out RIS phase-shift optimization using our proposed SFiD technique. First, we describe the RIS-assisted CF-mMIMO system and derive the achievable downlink sum-SE, while accounting for practical constraints, such as imperfect CSI and spatial correlation at the RIS. Next, we formulate the associated optimization problems for both continuous and discrete phase-shift cases and present their solutions using the proposed SFiD-C and SFiD-D techniques.
\vspace{-20pt}
\subsection{System Model}
\begin{figure}[htbp]
\includegraphics[width=1\linewidth,height=0.45\linewidth]{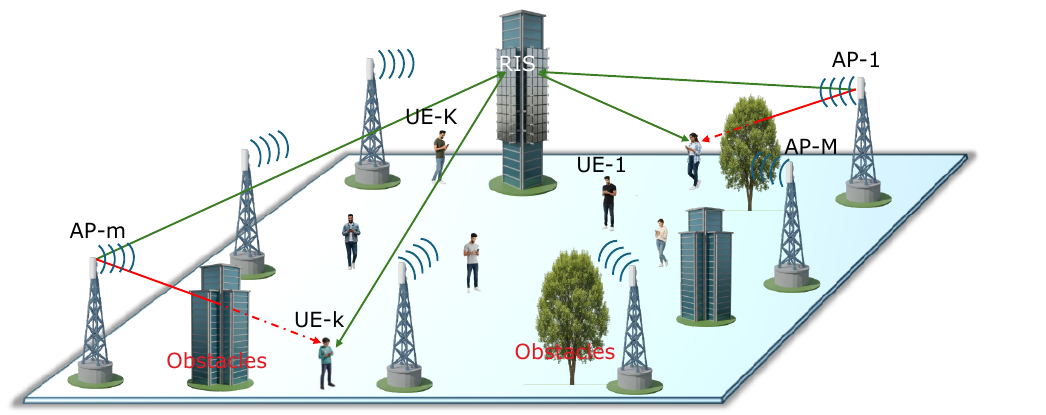}
\vspace{-15 pt}
\caption{\small RIS-assisted CF-mMIMO system model.} 
 \label{Fig2b:System model}
\end{figure}

We consider an RIS-assisted CF-mMIMO system comprising $M$ single-antenna APs connected to a CPU through fronthaul links, which collaboratively serve $K$ single-antenna UEs under time-division duplex operation using the same time-frequency resources, as shown in Fig. \ref{Fig2b:System model}. We consider a single RIS comprising $N$ reflecting elements. In this work, we adopt a spatially correlated Rayleigh fading channel model, where the AP-UE, AP-RIS, and RIS-UE channels are given by ${g}_{mk} = \beta_{mk}^{1/2} \bar{g}_{mk} \in \mathbb{C}$, 
$\mathbf{g}_{mr} = \mathbf{R}_{mr}^{1/2} \mathbf{\bar{g}}_{mr} \in \mathbb{C}^{N \times 1}$ and $ \mathbf{g}_{rk} = \mathbf{R}_{rk}^{1/2} \mathbf{\bar{g}}_{rk} \in \mathbb{C}^{N \times 1}$, ~respectively.
Here, $\bar{g}_{mk} \sim \mathcal{CN}(0, 1) $ and $\mathbf{\bar{g}}_{mr},\ \mathbf{\bar{g}}_{rk} \sim \mathcal{CN}(\textbf{0}, \textbf{I}_N)$ denote the small-scale fading components. The covariance matrices are defined as $\mathbf{R}_{mr} = \beta_{mr}A\mathbf{R}$ and $\mathbf{R}_{rk} = \beta_{rk}A\mathbf{R}$, with $\beta_{mk}$, $\beta_{mr}$, $\beta_{rk}$ representing the large-scale fading coefficients, $A\in \mathbb{R}$ denotes the RIS element area and $\mathbf{R} \in \mathbb{C}^{N \times N}$ characterizes the spatial correlation~\cite{CellFree_Chien2022RIS}. We denote the aggregate downlink channel impinging from the $m$th AP to the $k$th UE as
$h_{mk} = g_{mk} + \mathbf{g}_{rk}^H \mathbf{\Theta} \mathbf{g}_{mr}$. Here, $\mathbf{\Theta} \in \mathbb{C}^{N \times N}$ is the RIS phase-shift matrix given by $\mathbf{\Theta} = \text{diag}(e^{j\boldsymbol{\theta}})$, with $\boldsymbol{\theta}=[\theta_1, \theta_2, \cdots, \theta_N]$, where we have $\theta_i \in [0,2\pi)$ for continuous RIS phase-shift or $\theta_i \in \{0,\pi\}$ for discrete RIS phase-shift.

 \vspace{-5 pt}

In uplink training, all $K$ UEs transmit mutually orthogonal pilot signals. The APs then estimate the corresponding channel response using minimum mean-square error estimation~\cite{CellFree_Chien2022RIS}, given by $\hat{h}_{mk} = \{\sqrt{\tau_p \mathrm{p}_p} \bar{\beta}_{mk}\}/\{\tau_p \mathrm{p}_p \bar{\beta}_{mk} +1\}$. Here, \(\tau_p\) denotes the specific portion of the total coherence time \(\tau_c\) that is utilized for channel estimation, $\mathrm{p}_p$ denotes the normalized pilot power and the variance of the aggregated channel is given by $\bar{\beta}_{mk} = \beta_{mk} + \mbox{Tr}\left(\mathbf{\Theta} \mathbf{R}_{mr} \mathbf{\Theta}^H \mathbf{R}_{rk}\right)$.  

{In the downlink data transmission phase, each AP performs conjugate beamforming using statistical CSI to transmit the data~\cite{cellfree_Hien2017smallcell}. Let $q_k$ be the signal intended for UE $k$. Then signal transmitted from the $m$th AP is 
 \begin{align}
     x_m &= \sum_{k=1}^{K}  \frac{\sqrt{\mathrm{p}_d \eta_{mk}} \ \hat{h}^*_{mk}}{\sqrt{\mathbb{E}[|\hat{h}_{mk}|^2]}}  q_k = \sum_{k=1}^{K} \sqrt{\frac{\mathrm{p}_d \eta_{mk}}{\hat{\beta}_{mk}}}  \hat{h}^*_{mk}  q_k,
 \end{align}
where $\mathrm{p}_d$ denotes the maximum normalized power in the downlink and $\eta_{mk}$ is the power control coefficient of the $m$th AP to $k$th UE. The variance of the estimated channel is denoted as  $\hat{\beta}_{mk}$ and it is given by
$\hat{\beta}_{mk} = \{\tau_p \mathrm{p}_p \bar{\beta}_{mk}^2\}/\{\tau_p \mathrm{p}_p \bar{\beta}_{mk} +1\}$. Here, $\eta_{mk}$ is chosen such that $\mathbb{E}[|x_m|^2]\leq \mathrm{p}_d$, which leads to $\sum_{k=1}^{K}\!\eta_{mk}\leq~1$, where $\mathbb{E}[|q_{k}|^2] = 1$. The signal received at UE $k$ is 
\begin{align}
     y_k &= \sum_{m=1}^{M} h_{mk}x_m + z_k, \notag\\
     &= \sum_{m=1}^{M} \sum_{i=1}^{K} \sqrt{\frac{\mathrm{p}_d \eta_{mi}}{\hat{\beta}_{mi}}} h_{mk}\hat{h}^*_{mi}  q_{i} + z_k, \label{eq:cellfree_received_signal}
\end{align}
where $z_k \sim \mathcal{CN}(0, 1)$ denotes additive white Gaussion noise (AWGN) at the $k$th UE. We reformulate the received signal in \eqref{eq:cellfree_received_signal} using the use-and-then-forget technique of~\cite{cellfree_Hien2017smallcell} as
\begin{align}
     y_k &= \underbrace{\mathbb{E}\left[\sum_{m=1}^{M}\sqrt{\frac{\mathrm{p}_d \eta_{mk}}{\hat{\beta}_{mk}}} h_{mk}\hat{h}^*_{mk} \right] q_k}_{\text{Desired Signal}} \notag\\
     & \! \!+\! \underbrace{\left\{\!\sum_{m=1}^{M}\!\sqrt{\frac{\mathrm{p}_d \eta_{mk}}{\hat{\beta}_{mk}}} h_{mk}\hat{h}^*_{mk}\!-\!\mathbb{E}\!\left[\sum_{m=1}^{M}\!\sqrt{\frac{\mathrm{p}_d \eta_{mk}}{\hat{\beta}_{mk}}} h_{mk}\hat{h}^*_{mk} \right] \!\right\}q_k}_{\text{Beamforming gain uncertainty }} \notag \\
     & \ \ + \underbrace{\sum_{m=1}^{M} \sum_{\substack{i = 1 \\ i \neq k}}^{K} \sqrt{\frac{\mathrm{p}_d \eta_{mi}}{\hat{\beta}_{mi}}} h_{mk}\hat{h}^*_{mi}  q_{i}}_{\text{Interference}} \ + \underbrace{z_k}_{\text{AWGN}}.
     \label{eq:RX_signal}
\end{align}
The received signal in \eqref{eq:RX_signal} comprises the desired component, beamforming gain uncertainty, interference, and AWGN. A closed-form expression is derived for the achievable sum-SE as~\cite{CellFree_Chien2022RIS}
\begin{equation} \label{SumRate_formula}
  \mathcal{R}_{\text{sum}} (\boldsymbol{\theta}, \mathcal{C})= \sum_{k=1}^{K}\frac{\tau_c -\tau_p}{\tau_c}\log_{2}\big[1+\Delta_k(\boldsymbol{\theta}, \mathcal{C})\big],
\end{equation}
where $\Delta_k$ is SINR for $k$th UE in downlink and is given by
\begin{equation}
    \Delta_k (\boldsymbol{\theta}, \mathcal{C})= \frac{\mathrm{p}_d\left(\sum_{m=1}^{M} \sqrt{\eta_{mk}\hat{\beta}_{mk}}\right)^2}{\mathrm{p}_d \sum_{m=1}^{M} \sum_{k'=1}^{K} \eta_{mk'}\bar{\beta}_{mk}+1}.
\end{equation}
Here both $\hat{\beta}_{mk}$ and $\bar{\beta}_{mk}$ are functions of RIS phase-shift $\boldsymbol{\theta}$ and conditional information $\mathcal{C}$, where $\mathcal{C}$ is the set of large-scale fading coefficients. In the following subsection, we will utilize the proposed SFiD framework for optimizing the RIS phase-shift configurations in the CF-mMIMO system considered.
\vspace{-10 pt}
\subsection{Continuous RIS Phase-shift Optimization with SFiD-C}
\label{subsec:Cont_Phase_P1}
First, we formulate the optimization problem for the continuous RIS phase-shift with the objective of maximizing the achievable sum-SE. The continuous RIS phase-shift optimization Problem \textbf{P3} is  formulated as
\begin{subequations}\label{Problem_P1}
 \begin{align} \textbf{P3:}\quad
     \underset{\boldsymbol{\theta}}{\text{maximize}} \;\;& \mathcal{R}_{\text{sum}} (\boldsymbol{\theta}, \mathcal{C}),\label{P1_a} \\
     \text{subject to}\;\;  
     &\theta_{i} \in [0,2\pi), \;\;\; \forall i \in \{1, 2,\cdots, N\}, \label{P1_d} 
 \end{align} 
 \end{subequations} 
where, $\mathcal{C} = \{\beta_{mr}, \beta_{rk}, \beta_{mk}\},\forall m, k$. The constraint in~\eqref{P1_d} defines the legitimate range of phase-shifts for each RIS element. The optimization problem 
\textbf{P3} is non-convex and analytically challenging. To address these challenges, we employ the proposed SFiD-C framework for optimizing these continuous RIS phase-shifts in the CF-mMIMO system.

First, we use the conditional information consisting of $N_J$ large-scale fading coefficients corresponding to different UE and AP locations. 
Based on this conditional information, an initial continuous RIS phase-shift configuration is generated, which behaves similarly to a random phase initialization, since the model parameters are initially randomized.  Next, we compute the sum-SE (fitness value) for each conditional information using \eqref{SumRate_formula} and subsequently determine the corresponding normalized weight according to \eqref{eq:normalized_weight}. Then, we apply the forward diffusion process to the initialized continuous RIS phase-shift configuration and predict $\epsilon_w$. Using the normalized weight and the loss function defined in \eqref{eq:DM_ELBO}, we update the model parameters. This procedure is repeated iteratively until convergence is achieved. The detailed steps for SFiD-C training and inference are given in Algorithm~\ref{SFiD-DDPM Training} and Algorithm~\ref{SFiD-DDPM Inference}, respectively.

\begin{algorithm}
\caption{SFiD-C Training}
\label{SFiD-DDPM Training}
\begin{algorithmic}[1]
  \Statex \hspace*{-0.5 cm}\textbf{Input:} Number of epochs $N_{\text{ep}}$, conditional information $\mathcal{C}_j$ of given system, where $j\in \{1,\!\cdots\! N_{J}\}$.
  \Statex \hspace*{-0.5 cm}\textbf{Output:} A neural network trained with parameters $w$.
    \For{$n \gets 1$ \textbf{to} $N_{\text{ep}}$}
    \State Generate $\boldsymbol{\theta}_{ji}$ for given condition $\mathcal{C}_j$ by SFiD-C, for \Statex \hspace*{0.45 cm}  each  $i\in \{1,\!\cdots\!, N_{I}\}$ and $j\in \{1,\!\cdots\! N_{J}\}$.  
    \State Evaluate fitness value $r_{ji} = \mathcal{R}(\boldsymbol{\theta}_{ji}, \mathcal{C}_j)$.
  \State Find normalized weight \ $\tilde{\lambda}_{ji}$ \ for top-$k$ solutions as \Statex \hspace*{0.45 cm} given in \eqref{eq:normalized_weight} and add to dataset $\boldsymbol{\mathcal{D}}_\text{train}$.  
  \State Choose timestep $t \sim \text{Uniform}(1, 2, \dots T)$. 
  \State Get noisy RIS phase-shift
        $\boldsymbol{\theta}_{ji} $ at timestep $t$ by adding  \Statex \hspace*{0.45 cm}    Gaussion noise $\boldsymbol{\epsilon}$ to RIS phase-shift as given by \eqref{eq:forward_process1}.
    \State Predict \(\boldsymbol{\epsilon}_w\) by neural network using noisy RIS phase \Statex \hspace*{0.45 cm} shift
        $\boldsymbol{\theta}_{ji} $, $\mathcal{C}_j$ and $t$.
    \State Update the model using loss function defined in \eqref{eq:DM_ELBO} \Statex \hspace*{0.45 cm} with weight  $\Tilde{\lambda}_{ji}$. 
\EndFor

\end{algorithmic}
\end{algorithm}

 \begin{algorithm}
\caption{SFiD-C Inference}
\label{SFiD-DDPM Inference}
\begin{algorithmic}[1]
  \Statex \hspace*{-0.5 cm}\textbf{Input:} Conditional information $\mathcal{C}$ of given system and a trained model with parameters $w$.
  \Statex \hspace*{-0.5 cm}\textbf{Output:} Optimal $\boldsymbol{\theta}$ corresponding to $\mathcal{C}$.
    \State Generate random Gaussian $\boldsymbol{\theta}_T \sim \mathcal{N}(\boldsymbol{0, I})$.   
    \For{$t \gets T$ \textbf{to} $1$}
        \State Denoise $\boldsymbol{\theta}_t$ using \eqref{eq:Reverse_process_NN} with trained model to find  $\boldsymbol{\theta}_{t-1}$.  
    \EndFor 
\end{algorithmic}

\end{algorithm}

\vspace{-15pt}

 \subsection{Discrete RIS Phase-shift Optimization with SFiD-D}
 In the previous subsection, we studied the case where each RIS element can adjust its phase continuously. In practical deployments, however, hardware limitations impose finite-resolution control. To reduce circuit cost and implementation complexity, commercially realizable RISs typically support only a few quantized phase levels, and in many cases, each element is restricted to 1-bit discrete phase values for the simplest implementation. Hence, we consider discrete 1-bit RIS phase-shifts and formulate the optimization problem with the objective of maximizing the achievable sum-SE. The discrete 1-bit RIS phase-shift optimization Problem \textbf{P4} is  formulated as
\begin{subequations}\label{Problem_P2}
 \begin{align} \textbf{P4:}\quad
     \underset{\boldsymbol{\theta}^b}{\text{maximize}} \;\;& \mathcal{R}_{\text{sum}} (\boldsymbol{\theta}^b, \mathcal{C}),\label{P2_a} \\
     \text{subject to}\;\; 
     &\theta^b_{i} \in \{0,\pi\}, \;\;\; \forall i \in \{1, 2,\cdots, N\} \label{P2_d}, 
 \end{align} 
 \end{subequations} 
 where, $\mathcal{C} = \{\beta_{mr}, \beta_{rk}, \beta_{mk}\},\forall m, k$. The constraint in~\eqref{P2_d} restricts each RIS element to a $1$-bit discrete phase-shift\footnote{{In this work, 1-bit RIS phase-shifts are considered. The proposed SFiD-D and SFiD-D (BNN) techniques are readily extendable to multi-bit RIS phase-shifts configuration, which can be addressed in future work.}}, allowing values of either $0$ or $\pi$. Here, the optimization variable is discrete, resulting in a discontinuous and non-differentiable search space, which makes the problem inherently more complex and NP-hard. To tackle these challenges, we employ the proposed SFiD-D frameworks for efficiently optimizing the discrete RIS phase-shift configurations in the CF-mMIMO system.

We follow a similar procedure to that described in the previous subsection, with the key differences highlighted below. The optimized variable in SFiD-D corresponds to a 1-bit RIS phase shift, where the discrete phase values are constrained to 
 $0$ or $\pi$. In the forward diffusion process, instead of adding Gaussian noise, we apply a bitwise XOR operation using the binary perturbation $\boldsymbol{z}^b_t$. In addition, SFiD-D (BNN) employs a BNN, where the model weights are constrained to $1$-bit precision. Due to this binarization, a surrogate gradient function, as defined in \eqref{eq:Clipped_STE}, is utilized during back-propagation to enable effective model updates. The detailed procedures of SFiD-D and SFiD-D (BNN) training as well as inference are outlined in Algorithm~\ref{SFiD-BDPM Training} and Algorithm~\ref{SFiD-BDPM Inference}, respectively.

\begin{algorithm}
\caption{SFiD-D / SFiD-D (BNN) Training}
\label{SFiD-BDPM Training}
\begin{algorithmic}[1]
  \Statex \hspace*{-0.5 cm}\textbf{Input:} Number of epochs $N_{\text{ep}}$, conditional information $\mathcal{C}_j$ of given system, where $j\in \{1,\!\cdots\! N_{J}\}$.
  \Statex \hspace*{-0.5 cm}\textbf{Output:} A neural network trained with parameters $w$ ($w^b$)*.
    \For{$n \gets 1$ \textbf{to} $N_{\text{ep}}$}
    \State Generate $\boldsymbol{\theta}_{ji}^b$  for given condition $\mathcal{C}_j$ by model, for each \Statex \hspace*{0.51 cm}$i\in \{1,\!\cdots\!, N_{I}\}$ and $j\in \{1,\!\cdots\! N_{J}\}$. 
    \State Evaluate fitness value $r_{ji} = \mathcal{R}(\boldsymbol{\theta}_{ji}^b, \mathcal{C}_j)$.
    \State Find normalized weight \ $\tilde{\lambda}_{ji}$ \ for top-$k$ solutions as  \Statex \hspace*{0.51 cm}given in \eqref{eq:normalized_weight} and add to dataset $\boldsymbol{\mathcal{D}}_\text{train}$.  
  \State Choose timestep $t \sim \text{Uniform}(1, 2, \dots T)$. 
  \State Get perturbed RIS phase-shift
        $\boldsymbol{\theta}_{ji}^b $ at  $t$ by XORing  \Statex \hspace*{0.51 cm}perturbation $\boldsymbol{z}^b_t$ to RIS phase-shift as given by \eqref{eq:BDPM_noise_added}.
    \State Predict $\boldsymbol{z}^b_w$  by neural network using perturbed RIS  \Statex \hspace*{0.51 cm}phase-shift
        $\boldsymbol{\theta}_{ji}^b$, $\mathcal{C}_j\!$ and $t$ (Binarize weight value at\Statex \hspace*{0.51 cm}each step in neural network as given by \eqref{eq:binarization} )*.
    \State Update the model using loss function defined in \eqref{eq:BM_loss} \Statex \hspace*{0.42 cm} with weight  $\Tilde{\lambda}_{ji}$ (Update model using \eqref{eq:Clipped_STE})*.
\EndFor
\Statex \hspace*{-0.5 cm}\textbf{Note:} * for SFiD-D (BNN)

\end{algorithmic}

\end{algorithm}

\begin{algorithm}
\caption{SFiD-D / SFiD-D (BNN)   Inference}
\label{SFiD-BDPM Inference}
\begin{algorithmic}[1]
  \Statex \hspace*{-0.5 cm}\textbf{Input:} Conditional information $\mathcal{C}$ of given system and A trained model with parameters $w$ ($w^b$)*.
  \Statex \hspace*{-0.5 cm}\textbf{Output:} Optimal $\boldsymbol{\theta}^b$ corresponding to $\mathcal{C}$.
    \State Generate the binary variable $\boldsymbol{\theta}_T^b$ by sampling each element independently from a Bernoulli(0.5).   
    \For{$t \gets T$ \textbf{to} $1$}
        \State Denoise $\boldsymbol{\theta}_t^b$ using  trained model and find $\boldsymbol{\theta}_{t-1}^b$.  
    \EndFor 
\Statex \hspace*{-0.5 cm}\textbf{Note:} * for SFiD-D (BNN)
\end{algorithmic}

\end{algorithm}

\subsection{Complexity Analysis}

\vspace{-5pt}

{In this subsection, we analyze the computational complexity of the proposed method and compare it to both the GA and DRL-based deep deterministic policy gradient (DDPG) approach for continuous RIS phase-shift design, as well as to the DRL-based proximal policy optimization (PPO)  approach for discrete RIS phase-shift design. The computational complexity of the proposed SFiD-C method is expressed as $\mathcal{O}\big(N_{ep}N_JN_I(C_f + T C_n)\big)$ for the training stage, and $\mathcal{O}(T C_n)$ for the inference stage. Here $T$ denotes the number of diffusion time steps, $N_{ep}$ represents the number of training epochs, $N_JN_I$ is the number of training samples, $C_f$ corresponds to the complexity associated with the fitness evaluation and $C_n$ represents the computational complexity of the network architecture. The complexity of fitness function evaluation is given by $C_f \approx\mathcal{O}(MKN^3)$, where $M$ is number of APs, $K$ is number of UEs and $N$ is the number of RIS elements. In this work, we employ a U-Net architecture as the neural network backbone, whose complexity, quantified in floating-point operations, is given by 
$C_n = \mathcal{O}(ND)$~\cite{In_012_CDDPM_letafati_2025}. The computational complexity of the GA depends on the number of generations $(G)$, the population size $(P)$ and the complexity of the fitness function evaluation $(C_f)$, and  expressed as $\mathcal{O}(GPC_f)$. The computational complexity of the DDPG algorithm in the training stage depends on the number of episodes $(E)$, the length of each episode $(L)$, the complexity of the actor–critic networks $(C_{act}-C_{crit})$ and the cost of reward evaluation $(C_f)$. The overall complexity is therefore given by $\mathcal{O}\big(EL(C_{act}+C_{crit}+C_f)\big)$. In the evaluation stage, only forward passes through the actor network are required, leading to a reduced complexity of $\mathcal{O}\big(E_e(C_{act}+C_f)\big)$, where $E_e$ is evaluation steps.}

Note that in SFiD-C, although training is computationally expensive due to per-epoch fitness evaluation and the large number of iterations required for convergence, it leads to improved performance during the inference stage. Moreover, this additional cost does not affect inference, where the fitness function is no longer required, making inference computationally more efficient than the GA and the DDPG. In the GA, convergence to near-optimal solutions generally requires high values of $G$ and $P$, resulting in a high computational cost. Moreover, when fitness function evaluation is computationally intensive, the overall complexity further increases, making the GA less efficient for large-scale problems. In the DDPG, achieving stable convergence typically requires a large number of episodes and updates, along with repeated reward evaluations, which substantially increase the computational cost. Overall, the GA is computationally expensive, and although the DDPG reduces the complexity, the proposed SFiD-C method significantly lowers the computational cost at a performance comparable to that of the GA.

\begin{figure}[htbp]
\vspace{-10pt}
\includegraphics[width=0.9\linewidth,height=0.7\linewidth]{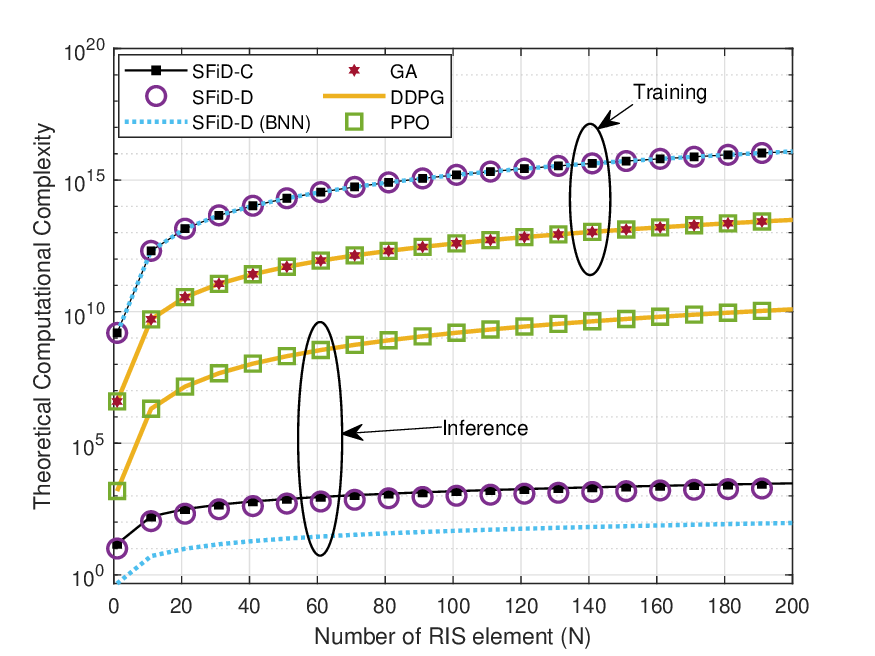} 
\vspace{-8pt}
\caption{\small Computational complexity of different algorithms for $M =64$, $K = 12$, $G = 100$, $P= 50$, $E = 100$, $L=50$, $E_e = 2$ and $D=3$.}  \label{Fig3:Final_complex}
\end{figure}

We now examine the computational complexity of SFiD-D and SFiD-D (BNN) for discrete phase-shift optimization. As their overall procedures closely follow that of SFiD-C with a few key modifications, the resultant complexity is comparable, while being reducible through the use of bitwise operations. In SFiD-D and SFiD-D (BNN), the binary input is XORed with binary perturbations instead of adding Gaussian noise to real-valued inputs. Both elementwise addition and XOR operations scale linearly with the number of elements. In practice, XOR operations are more efficient when applied to bit-packed binary data, as they enable parallel processing of multiple elements per instruction, unlike floating-point additions that operate element-wise. This results in a lower constant computational cost, although the performance attained for large tensors remains limited by memory. Additionally, SFiD-D (BNN) employs binarization and implementing convolutions via XNOR and popcount operations provide substantial computational gains, yielding up to $58\times$ faster performance than SFiD-D~\cite{Mot_002_BNN_survey_2023, BNN_new_2024}. For PPO, the training cost additionally depends on the number of policy update epochs $E_k$, since the clipped surrogate objective is optimized for multiple passes over the collected trajectory. Hence, the PPO incurs $\mathcal{O}\big(EL(E_kC_{act}+E_kC_{crit}+C_f)\big)$ complexity during training, and $\mathcal{O}\big(E_e(C_{act}+C_f)\big)$ during evaluation.

\begin{figure*}[htbp]
	\centering
	\begin{subfigure}[b]{0.32\linewidth}
    \centering
		\includegraphics[width=1\linewidth,height=0.9\linewidth]{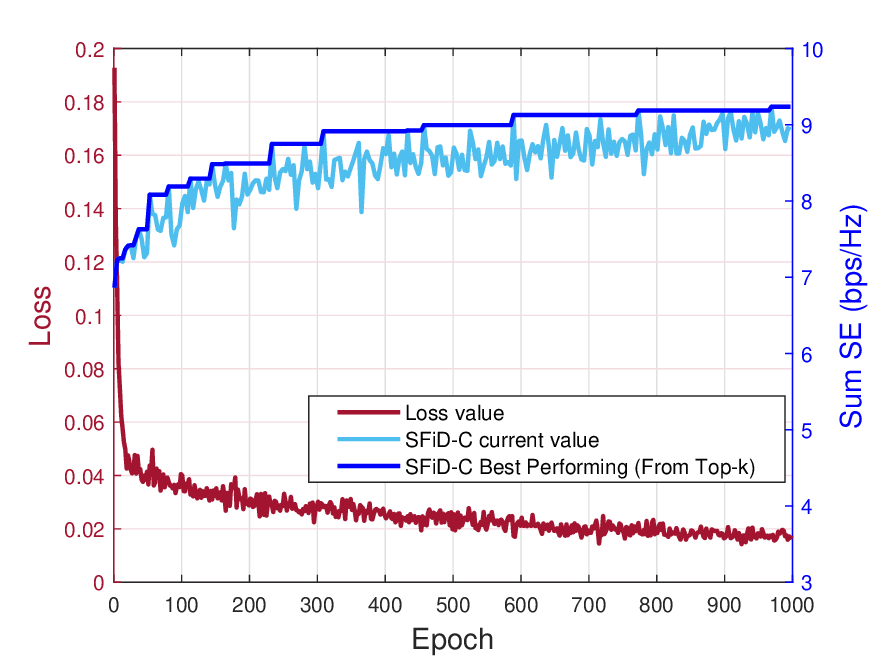}
            \caption{}
		\label{fig:5_1}
	\end{subfigure}\hfill
	\begin{subfigure}[b]{0.32\linewidth}
    
    \includegraphics[width=1.1\linewidth,height=0.9\linewidth]{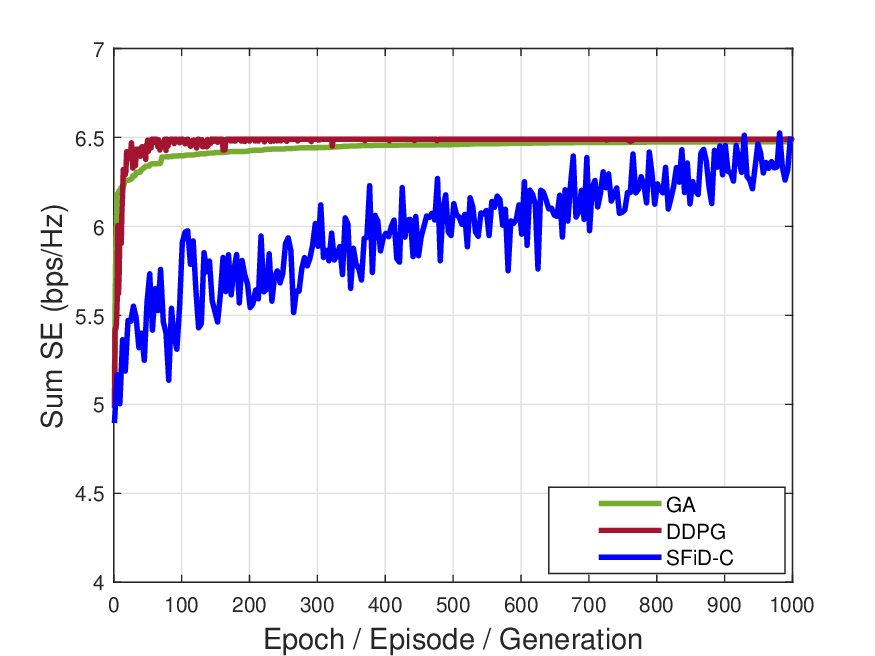}
        \caption{}
	\label{fig:5_2}
	\end{subfigure}\hfill
	\begin{subfigure}[b]{0.32\linewidth}
    
	\includegraphics[width=1.1\linewidth,height=0.9\linewidth]{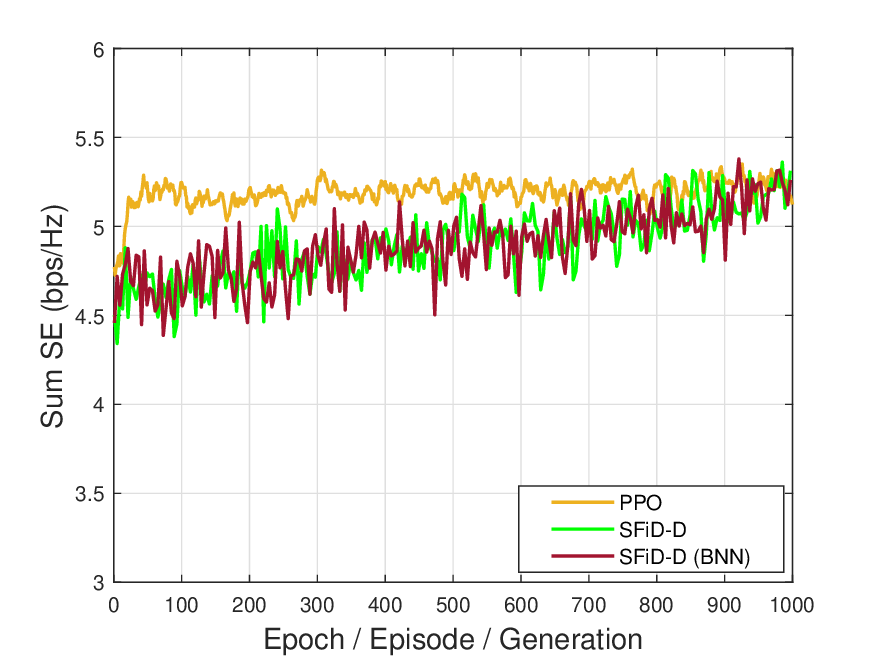}
        \caption{}
	\label{fig:5_3}
	\end{subfigure}\hfill
    \vspace{-10pt}
	\caption{\small a) Training convergence of the proposed SFiD-C method, b) Convergence comparison on a test sample for continuous RIS phase-shift optimization, c) Convergence comparison on a test sample for discrete RIS phase-shift optimization. \\[-30pt]} 
\end{figure*} 

Overall, SFiD-D requires a computational complexity comparable to SFiD-C, which can be reduced through the use of bitwise operations. Furthermore, in SFiD-D (BNN), the inference-stage complexity is further reduced, since real-valued additions and multiplications are replaced by binary operations.
Moreover, for discrete action spaces, the PPO applies a softmax normalization before the argmax action selection, incurring additional computational overhead that scales with the number of possible actions. In the evaluation stage, the PPO still requires reward evaluation, which increases latency, when the fitness computation itself is complex. In summary, SFiD-D and  SFiD-D (BNN) incur a lower computational cost than PPO during inference, highlighting their efficiency and suitability for practical deployment.

The above analysis of computational complexity is characterized in Fig.~\ref{Fig3:Final_complex}. The figure illustrates the computational complexity of various algorithms - ~SFiD-C, SFiD-D, SFiD-D (BNN), the GA, the DDPG and the PPO - during both the training and inference stages. They are evaluated for different values of $N$, representing the number of RIS elements. The figure illustrates that the proposed algorithms involve higher computational cost during training but require significantly lower computation during the inference phase, making them well-suited for real-time applications. Since the inference stage eliminates the need for computationally intensive fitness evaluations, the overall complexity is substantially reduced.
Thus, the proposed methods, SFiD-C and SFiD-D, are computationally more efficient during the inference stage than the DDPG and the PPO.


\vspace{-5 pt}

\section{Numerical Results and Analysis}

In this section, we describe the experimental setup and present numerical results for validating the performance of our proposed SFiD-C and SFiD-D algorithms for RIS phase-shift optimization in a CF-mMIMO system.

\vspace{-10 pt}

\subsection{Experimental Setup}
We employ a modified U-Net architecture. The network consists of an encoder having a single down-sampling layer, a middle layer, and a decoder with a single up-sampling layer. A self-attention layer is incorporated for enhancing the overall performance. The model embeds time-dependent information using a dimension of $64$ and incorporates conditional embeddings with a dimension of $128$. A dimensionality of $64$ and $128$ is chosen to strike a balance between computational efficiency and performance in the diffusion model.
For allowing the model to adapt to different scenarios, it concatenates conditional information with the input channels, enabling it to learn and generate outputs that align with specific conditions. Additionally, it integrates temporal and contextual data for capturing dependencies throughout the learning process. We use the AdamW~\cite{AdamW_2024} optimizer with a multi-step learning rate of $1\times 10^{-4}$ to ensure stable convergence. Furthermore, we apply a weight decay factor of $0.25$ to regularize the training process and mitigate overfitting. Additionally, to ensure stable training, an exponential moving average strategy is employed for updating the model parameters during training~\cite{EMA_2020}. In our simulations, we consider a $1\,\text{km} \times 1\,\text{km}$ coverage area, where the APs and UEs are randomly positioned. The distance-dependent large-scale path loss is modeled using a three-slope path loss model~\cite{cellfree_Hien2017smallcell}. We configure the network with $M=64$ APs, serving 
$K=12$ UEs and an RIS consisting of $N=64$ elements. We conducted all experiments on a Windows 11 system with a 10th Gen Intel(R) Core(TM) i7-10700 CPU at 2.90 GHz and 32 GB RAM.

\vspace{-10pt}
\subsection{Convergence Analysis}
Fig.~\ref{fig:5_1} presents the sum-SE of training samples and the loss evolution versus the training epochs for the proposed SFiD method. We observed that the algorithm conceived begins to converge at approximately 1000 training epochs. This behavior indicates that the diffusion-based optimization process gradually learns an effective mapping from noisy latent representations to optimized RIS phase configurations. Furthermore, the decaying loss trend aligns well with the improvement in sum-SE, confirming that the model is consistently moving toward better solutions, rather than overfitting to intermediate states. Overall, these observations highlight both the stability and learning efficiency of the proposed SFiD method during training.

{Fig.~\ref{fig:5_2} and Fig.~\ref{fig:5_3} characterize our convergence analysis for continuous and discrete RIS phase-shift optimization, respectively. For continuous RIS phase-shift optimization, the proposed SFiD-C is compared to both GA and DDPG, whereas for discrete RIS phase-shift optimization, our proposed SFiD-D and SFiD-D (BNN) schemes are compared with PPO on the test sample. Note that, in our approach, the model is trained on a training set that is different from the test samples, whereas the baseline methods perform both training and testing on the same test samples\footnote{When the training and testing samples follow different distributions, PPO exhibits good generalization and stable performance than DDPG, whereas DDPG often suffers from performance degradation due to its off-policy and deterministic nature~\cite{PPO_1_2021}. So, we evaluate DDPG using identical training and testing samples, while PPO is evaluated under different training and testing sample distributions.}. In Fig.~\ref{fig:5_2}, the GA converges slowly, approaching the maximum after about $1000$ generations. By contrast, the DDPG algorithm converges rapidly, approaching peak performance within $20$ episodes and saturating beyond $300$ episodes. Our proposed SFiD-C converges more slowly than both the GA and DDPG, reaching its maximum performance after roughly $1000$ epochs. In Fig.~\ref{fig:5_3}, the convergence behaviour of SFiD-D and SFiD-D (BNN) is similar and takes more iterations than PPO. PPO converges rapidly, approaching its peak performance within 20 episodes. These results indicate that, during the training phase, the proposed approach requires more iterations to achieve convergence. However, it shines during the inference stage, as we will discuss in the following subsection, justifying this additional training effort. Training in the proposed method is performed offline in both the continuous and discrete cases. Therefore, increasing the number of epochs does not impact inference time, but beneficially enhances training stability and increases the likelihood of achieving improved performance.}

\vspace{-10pt}

\subsection{Performance of SFiD-C for Continuous RIS Phase-shift Optimization}

Fig.~\ref{Fig6:SFiD_DDPM_Performance} and~\ref{Fig7:SFiD_DDPM_Time} illustrate the effectiveness of the proposed SFiD-C technique by comparing its performance to three approaches: random allocation, the GA, and the DDPG for continuous RIS phase-shift optimization. For DDPG, we evaluate two cases, one in which the training and testing samples are identical (DDPG-I) and another in which they are different (DDPG-D). This comparison highlights that the proposed SFiD-C consistently outperforms both conventional heuristics as well as advanced learning-based techniques, thereby validating its robustness and efficiency.
Fig.~\ref{Fig6:SFiD_DDPM_Performance} shows that the proposed SFiD-C approach consistently outperforms random phase configurations, underscoring the critical role of optimized phase-shift design. The proposed method demonstrates superior performance compared to DDPG-I and achieves performance similar to that of the GA and DDPG-D, highlighting its effectiveness in RIS phase-shift optimization. We observe that the DDPG performs well when the training and testing samples are identical (DDPG-I), since the learned policy directly exploits the underlying data distribution. However, its performance degrades noticeably under distribution shift (DDPG-D) due to its limited generalization beyond the training data. Furthermore, while the DDPG-I achieves superior performance compared to the DDPG-D, this improvement comes at the cost of higher latency, as the model must be trained on samples available only at runtime, rather than offline. 

\begin{figure}[htbp]
	\centering
	\begin{subfigure}[b]{0.5\linewidth}
    \centering
		\hspace{-15pt}\includegraphics[width=1.09\linewidth,height=0.9\linewidth]{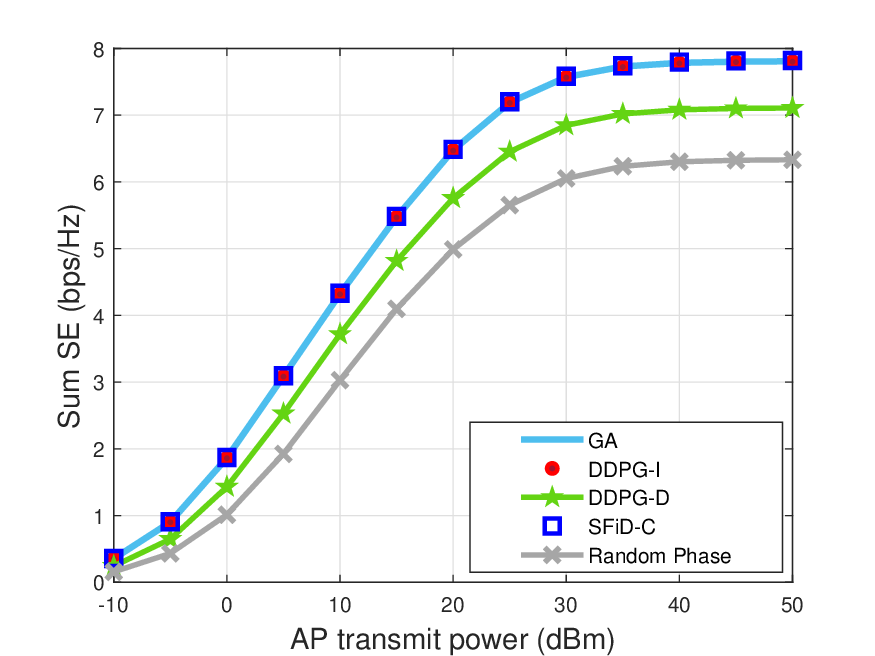}
            \caption{}
		\label{Fig6:SFiD_DDPM_Performance}
	\end{subfigure}\hfill
	\begin{subfigure}[b]{0.5\linewidth}
    
    \includegraphics[width=1.09\linewidth,height=0.9\linewidth]{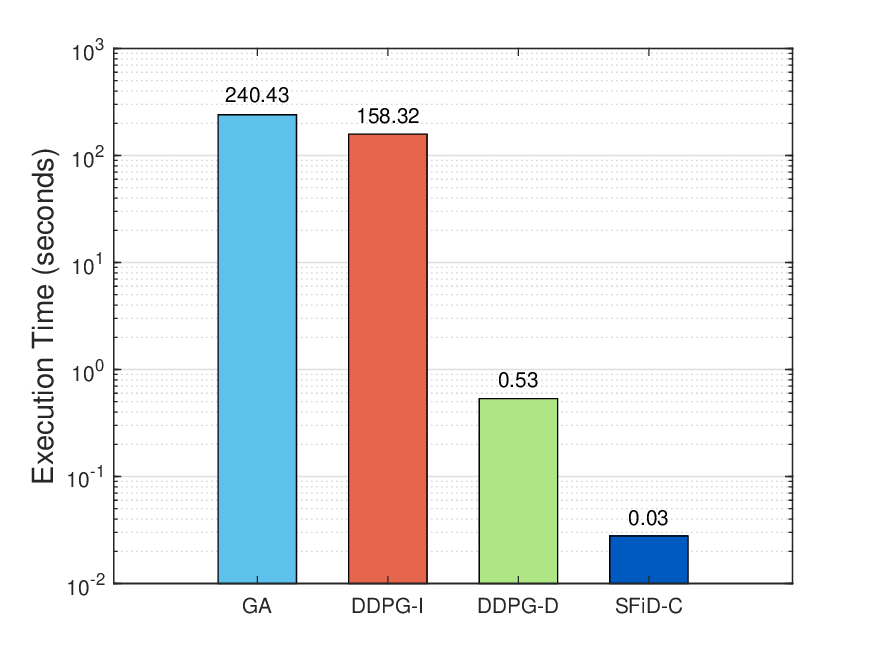}
        \caption{}
	\label{Fig7:SFiD_DDPM_Time}
	\end{subfigure}\hfill
	\hfill
    \vspace{-10pt}
	\caption{\small Continuous RIS phase-shift optimization: a) Sum-rate performance comparison for different algorithms, b) Computational run time comparison for different algorithms.} 
\end{figure}

The computational efficiency of the proposed SFiD-C was evaluated and compared to both the GA and the DDPG in Fig.~\ref{Fig7:SFiD_DDPM_Time}. For the GA, the computation time required for RIS phase-shift optimization is $240.43$ seconds, whereas the proposed SFiD-C method reduced it to only $0.03$ seconds, representing a reduction to approximately $0.01\%$. Similarly, DDPG-I and DDPG-D, which require $158.32$ seconds and $0.53$ seconds, respectively, SFiD-C reduces the computation time to $0.03$ seconds, corresponding to only $0.01\%$ and $5.66\%$ of their runtimes.  These results clearly demonstrate that the SFiD-C significantly reduces the runtime while maintaining or surpassing the performance of existing methods, making it an efficient practical solution for real-time RIS phase-shift optimization in CF-mMIMO systems.

\subsection{Performance of SFiD-D for Discrete RIS Phase-shift Optimization}

Both Fig.~\ref{Fig8:SFiD_BDPM_BN_Performance} and \ref{Fig9:SFiD_BDPM_BN_Time} demonstrate the efficiency of the SFiD-D and SFiD-D (BNN) techniques for discrete RIS phase-shift configurations in RIS-assisted CF-mMIMO systems. First, we compare these results with the optimized continuous RIS phase-shift configuration seen in Fig.~\ref{Fig8:SFiD_BDPM_BN_Performance} using our SFiD-C technique. As expected, a performance degradation is observed when moving from continuous to discrete phase-shifts. Then, we compare the performance of SFiD-D and SFiD-D (BNN) to both random discrete phase-shift allocation and to the PPO baseline. As shown in Fig.~\ref{Fig8:SFiD_BDPM_BN_Performance}, the proposed SFiD-D and SFiD-D (BNN) outperform the random allocation baseline and marginally outperform the PPO method, while Fig.~\ref{Fig9:SFiD_BDPM_BN_Time} demonstrates the notable computational efficiency of our proposed methods. Fig.~\ref{Fig9:SFiD_BDPM_BN_Time} compares the computational efficiency of the proposed methods, SFiD-D and SFiD-D (BNN), to the PPO approach. The PPO method requires approximately $0.83$ seconds for inference, whereas our SFiD-D and SFiD-D (BNN) achieve much faster execution, requiring only $0.03$ seconds and $0.017$ seconds, respectively. SFiD-D (BNN) achieves lower inference time by employing a BNN. However, the achieved latency is still higher than the theoretical expectation. This gap arises because the current implementation does not yet leverage XNOR-Popcount operations. To fully realize the primary advantage of BNNs (i.e., substantial reduction in inference time), specialized libraries or hardware support are required to convert standard convolution operations (Conv2D) into their optimized XNOR-Popcount form.

\begin{figure}[htbp]
	\centering
	\begin{subfigure}[b]{0.5\linewidth}
    \centering
		\hspace{-15pt}\includegraphics[width=1.09\linewidth,height=0.9\linewidth]{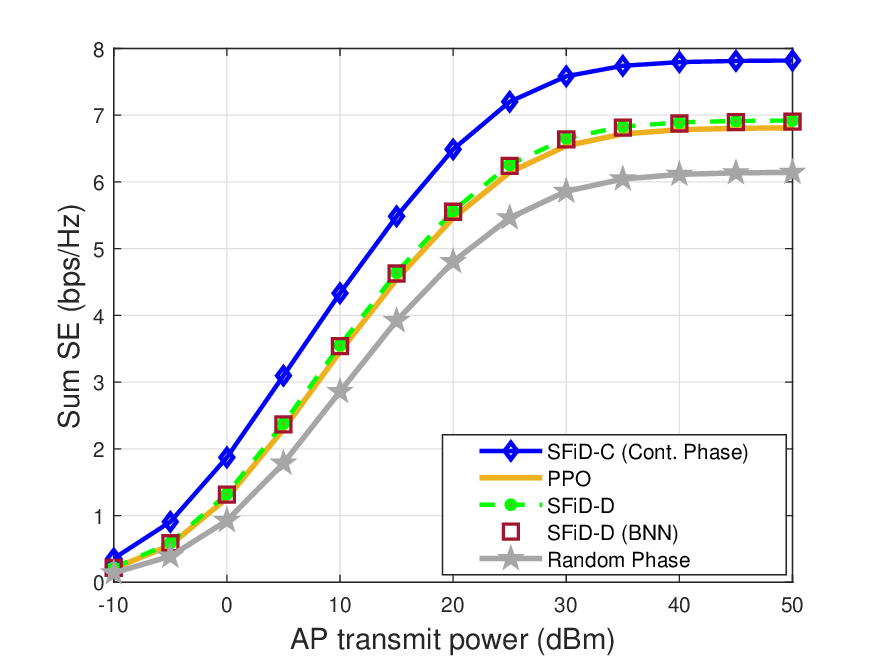}  
            \caption{}
		\label{Fig8:SFiD_BDPM_BN_Performance}
	\end{subfigure}\hfill
	\begin{subfigure}[b]{0.5\linewidth}
    
    \includegraphics[width=1.09\linewidth,height=0.9\linewidth]{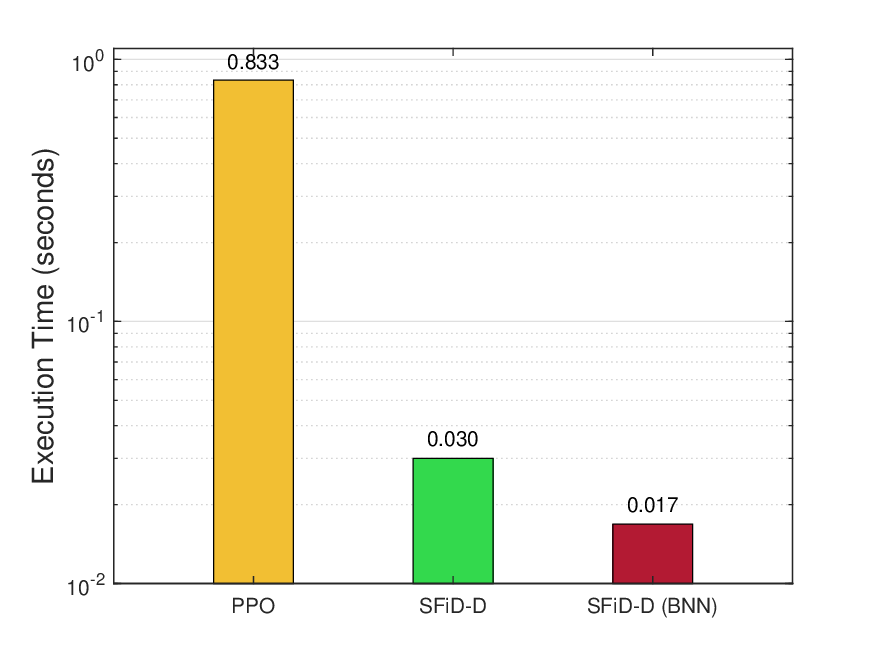}
        \caption{}
	\label{Fig9:SFiD_BDPM_BN_Time}
	\end{subfigure}\hfill
	\hfill
    \vspace{-5pt}
	\caption{\small Discrete (1-bit) RIS phase-shift optimization: a) Sum-rate performance comparison for different algorithms, b) Computational run time comparison for different algorithms.} 
\end{figure}

Furthermore, we provide a summary of the storage requirements of our different proposed techniques. The SFiD-C and SFiD-D models require approximately the same memory of about $1.2$ MB. In the SFiD-D (BNN) model, when all weights and biases are binarized (BNN-style) and packed into 32-bit words,  the storage decreases to $0.06$ MB, resulting in about a $20\times$ memory reduction. Note that this reduction is lower than the theoretical $32\times$ factor, since a portion of the model parameters (e.g., normalization statistics, embedding indices, tensor shapes/headers, and other supporting information) must still be stored in floating-point format and therefore cannot be fully binarized. Moreover, this reduced memory footprint leads to a lower model loading time, which further contributes to a reduction in overall inference time.

\section{Conclusions}

A self-supervised fitness-driven diffusion (SFiD) technique was conceived for solving complex optimization problems, addressing the limitations of conventional optimization techniques and AI-based methods that rely on labeled data or exhibit limited generalization. Specifically, we developed SFiD-C for continuous variable optimization and SFiD-D along with SFiD-D (BNN) for discrete variable optimization. We validated the convergence of the proposed method to the global optimum in the context of analytical convex and non-convex optimization problems. By considering continuous and 1-bit binary RIS phase-shift optimization in CF-mMIMO systems under practical constraints, we demonstrated the efficiency of our proposed methods. The results demonstrate that SFiD-C, SFiD-D and SFiD-D (BNN) achieve comparable or superior performance to benchmark methods, while significantly improving computational efficiency. This highlights their efficiency in handling both continuous and discrete decision variables in complex optimization problems. Overall, this treatise establishes a promising framework for self-generated expert training data based optimization, paving the way for future extensions to more diverse large-scale, and real-world optimization problems in wireless communication systems.

\bibliographystyle{IEEEtran}
\vspace{-10pt}
\bibliography{cite}

\end{document}